\documentclass[prd,showpacs,twocolumn,superscriptaddress,nofootinbib]{revtex4-2}
\usepackage{mathrsfs}
\usepackage{yfonts}
\usepackage{tipa}
\usepackage{bm}
\usepackage{bbm}
\usepackage{amsmath}
\usepackage{amssymb}
\usepackage{amsthm}
\usepackage{amsfonts}
\usepackage{mathtools}
\usepackage{latexsym}
\usepackage{relsize}
\usepackage[english]{babel}
\usepackage{booktabs}
\usepackage[utf8]{inputenc}
\usepackage{tikz}
\usepackage{array}
\usepackage{hyperref, color}
\usepackage{comment, multirow}
\usetikzlibrary{shapes.geometric, arrows.meta, positioning}

\newcommand{\bea}{\begin{eqnarray}}
	\newcommand{\eea}{\end{eqnarray}}
\newcommand{\be}{\begin{equation}}
	\newcommand{\ee}{\end{equation}}
\newcommand{\ba}{\begin{align}}
	\newcommand{\ea}{\end{align}}

\begin{document}
	\title{Constraining Early Dark Energy cosmological models with Big Bang Nucleosynthesis}
	\author{Teodora Maria Matei}
	\email{teodora.matei@acad-cj.ro}
	\affiliation{Astronomical Institute of Romanian Academy, Cluj-Napoca Branch, 19 Cire\c silor Street, 400487 Cluj-Napoca, Romania}
	\affiliation{Department of Physics, Babe\c s-Bolyai University, Kog\u alniceanu Street, Cluj-Napoca, 400084, Romania,}
	
	\author{Cristian Croitoru}
	\email{croitoru.lu.cristian@student.utcluj.ro}
	\affiliation{Faculty of Automation and Computer Science, Technical University of Cluj-Napoca, George Baritiu Street, 400027 Cluj-Napoca, Romania,}
	
	\author{Tiberiu Harko}
	\email{tiberiu.harko@aira.astro.ro}
	\affiliation{Department of Physics, Babe\c s-Bolyai University, Kog\u alniceanu Street, Cluj-Napoca, 400084, Romania,}
	\affiliation{Astronomical Institute of Romanian Academy, Cluj-Napoca Branch, 19 Cire\c silor Street, 400487 Cluj-Napoca, Romania}

	\begin{abstract}
		The recent cosmological picture contains a significant tension indicating that our standard $\Lambda$CDM picture may be incomplete.  Early Dark Energy models can alleviate the Hubble tension,  by assuming an early acceleration that could explain the divergence between the early and late-time cosmological data. We investigate the implications of Early Dark Energy models on the Big Bang Nucleosynthesis processes by considering several cosmological models, including a model assuming a simple cosmological constant, alongside with varying equations of state  dark energy models. We construct a simulator through a nested sampling algorithm, with the help of which we estimate the upper bounds for model parameters, and determine the maximum allowable dark energy density contribution during the radiation-dominated era. Our results are obtained through the \href{https://github.com/croi900/eden}{eden} program. We  show that for a linear or polytropic equation of state, the dark energy density is constrained to less than $10^{-13}$ MeV$^4$ and $10^{-5}$ MeV$^4$, respectively, at the 95\% confidence level. Furthermore, we identify a temperature-dependent equation of state of dark energy as the most physically compelling framework, which remains consistent with primordial abundances for coupling parameters $\lesssim 10^{-2}$. This model successfully allows for high-temperature deviations from the standard $\Lambda$CDM expansion history, while rapidly diluting to obtain standard general relativistic results in the weak freeze-out era.
	\end{abstract}
	
	\maketitle
	\tableofcontents
	
	\section{Introduction}
	
	The standard cosmological model $\Lambda$CDM, relies on the presence of a cosmological constant denoted $\Lambda$, which drives the late-time accelerated expansion of the Universe, a phenomenon firstly confirmed by Type Ia supernovae observations \cite{art1, art2}, and more recently by \cite{Planck2018, art4}. In the present-day Universe, observations constrain this constant to $\Lambda \approx 10^{-56}\text{ cm}^{-2}$ \cite{Planck2018}. 

However, recently, the $\Lambda$CDM model is facing significant challenges from the new BAO measurements of the first data release of the Dark Energy Spectroscopic Instrument (DESI), with the second data release (DR2) raising even more questions about the validity of the standard cosmological paradigm  \cite{D1,DESI}. When DESI DR2 BAO data are combined with the CMB data from Planck, the Atacama Cosmology Telescope (ACT) \cite{ACT}, and Supernova datasets (whether from Union3, Pantheon Plus, or DESY5),  deviations from the $\Lambda$CDM paradigm become apparent, leading to the necessity of considering Dynamical Dark Energy (DDE) models \cite{N1}. For example, when using the CMB+DESI+DESY5 data,  the Barboza-Alcaniz (BA) model gives $w_0 = -0.785 \pm  0.047$ and $w_a = -0.43^{+0.10}_{ -0.09}$, a result which notably deviates from the $\Lambda$CDM prediction, and provides convincing evidence for DDE at the 4.2$\sigma$ level \cite{N1}. For a review of the recent results and challenges of the Dynamical Dark Energy models see \cite{N2}.  

The compelling evidence for Dynamical Dark Energy models naturally raises the question of the presence, and relevance of the dark energy in the early, and very early Universe. Did, for example, the value of the cosmological constant change in time?  Even though late-time constraints for the value of the cosmological constant have been obtained from observations, its value in the early Universe is not yet known. Hence the natural question arises of what is the maximum allowed effective value of $\Lambda$ during the first stages of the cosmological evolution. 

Therefore,  the study of the influence of the cosmological constant $\Lambda$ on the dynamics of the early Universe, as well as the consideration of the dynamically equivalent candidate models, known as Early Dark Energy (EDE) models may lead to a new perspective on the early Universe. 

The introduction of an EDE component is primarily motivated by the Hubble tension, i.e. the discrepancy between local $H_0$ measurements and those inferred from the early Universe \cite{art6, art7}. Technically, EDE models postulate the existence of a dynamical component that contributes a non-negligible fraction to the total energy density during the early radiation-dominated era, which can raise the $H_0$ parameter in order to alleviate the tension, initially handled through quintessential scalar fields or high-energy phase transitions \cite{Wetterich2004, Doran2006, Pettorino2013}. 

This field of research was developed by \cite{ede1, ede2}, in which axion-like potentials were proposed, aiming to resolve the Hubble tension. Some other early works included Acoustic Dark Energy which treated acoustic oscillations in the dark component \cite{ade}, New Dark Energy which accounted for a phase transition before recombination \cite{nede} and Rock 'n' Roll oscillatory and rolling scalar field EDE \cite{rnr}. For recent reviews of EDE framework see \cite{rev1, rev2}. In the context of modified gravity, dynamical dark energy was studied in $f(R)$ theories \cite{fR}, $f(T)$ teleparallel gravity \cite{fT}, and Early Modified Gravity \cite{EMG1, EMG2} in which non-minimally coupled scalar fields affect the early Universe's dynamics prior to recombination.
	
	High-precision results from recent ACT DR6 and DESI DR2 \cite{ACT, DESI} suggest that in order for the early dark energy model parameters to be compatible with CMB/LSS requirements, the dark component is restricted to an energy fraction $f_{EDE} < 0.016$. The DESI DR2 observations favour a late-time dynamical dark energy framework, recently studied through Critically Emergent Dark Energy \cite{cede} and in the early Universe, cosmic birefringence acts as a parity-violating signature of the dark sector \cite{Bir1, Bir2}. Other works that constrain EDE components by both DESI and ACT datasets, are mentioned \cite{Lodha2025, Chaussidon2025, Qu2025, Wang2025}. 
	
	Besides the CMB data constraints, Big Bang Nucleosynthesis (BBN) can impose upper limits by measuring weak freeze-out deviations from a modified Hubble expansion rate, altering the primordial nuclei formation. 

Big Bang Nucleosynthesis is a theoretical framework that predicts light element formation in the early times of the Universe, such as Hydrogen, Deuterium, Helium and Lithium (see \cite{BBN1, bbn-1, bbn-2, bbn-3, bbn-4, BBN2, Cooke2018, Aver2021, BBN3, BBN4} for early works and comprehensive review articles), that result in abundance ratios $Y_p = $ He$^4$/H or D/H, which are in agreement with recent astrophysical measurements \cite{pdg}, up to the theoretical prediction of Li$^7$/H, which remains in high disagreement with observations \cite{BBN5}.
	
	Any deviation from the GR limit can be restricted through the abundance ratios provided through nucleosynthesis, so that BBN can probe frameworks that depart from $\Lambda$CDM model. Within such extensions of the standard framework, scalar-tensor theories \cite{Santiago1996, Coc2006}, varying gravitational constant scenarios \cite{Alvey2020}, $f(R, T)$ gravity \cite{Sahoo}, $f(T)$ teleparallel gravity and generalized $f(T,T)$ models \cite{Capozziello2017, Bhattacharjee2024, Cruz2025}, modified Gauss-Bonnet gravity \cite{Kusakabe2016}, and bimetric gravity \cite{Hogas2021}, were studied and constrained with the help of nucleosynthesis yields. Recently, these constraints have been extended to more complex geometric and high-energy frameworks, such as \cite{Matei2025a, Matei2025b} which explored the phenomenological implications of Weyl boundary in the early Universe and space-time noncommutativity, and also \cite{Cook2024}, which proved that modified gravitational couplings during the MeV era must remain highly suppressed in order to predict primordial deuterium and helium formation.
	
	Beyond modified gravity, BBN also probes high-energy physics and the dark sector, for example neutrino-extended Effective Field Theories \cite{Braat2026} and electron neutrino-sterile neutrino oscillations \cite{Kirilova2024} were recently constrained through nuclear abundances. In the dark matter sector, both resonantly-enhanced \cite{Depta2024} and sub-GeV hadronic annihilations \cite{Omar2026}, as well as on the lifetimes of heavy QCD axions \cite{Axion2025} have been restricted by the precise astrophysical measurements of the helium and deuterium mass fractions.

By studying the EDE component in the MeV scale, the increase of the Hubble rate determines an overproduction of helium and deuterium, such that BBN serves as a hard boundary on the permissible energy fraction of the dark sector. Recent analyses restricted these bounds, for instance EDE-like scalar fields were limited based on combined BBN and sound-horizon-independent CMB lensing in \cite{Kable2024}. The direct impact of an EDE density scaling as in \cite{McKeen2024}, reconfirms that the EDE fraction must remain highly subdominant to preserve concordance with observations. A principal component analysis was applied in \cite{Cook2025} to impose model-independent constraints on any dark energy history during BBN. Furthermore, EDE was constrained alongside varying electron mass scenarios by using the latest DESI DR2, ACT DR6, and BBN datasets, finding tight upper limits on the allowed dark energy fraction \cite{Seto2025}. Together, these works require that any viable EDE solution to late-time cosmological tensions must either remain subdominant during the first three minutes or exist strictly below the energy thresholds provided by primordial nucleosynthesis.
	
	In order to simulate the processes that may have taken place in the early Universe and compare theory to data, various packages were constructed such as \texttt{CLASS} \cite{Blas2011} or \texttt{CAMB} \cite{Lewis2000}, which are widely used as Boltzmann solvers to compute CMB anisotropies and large-scale structure observables. When evaluating primordial abundances, these solvers are typically interfaced with dedicated Big Bang Nucleosynthesis networks like \texttt{PArthENoPE} \cite{Pisanti2008} or \texttt{AlterBBN} \cite{Arbey2020}. For parameter estimation, these codes are integrated into Bayesian inference frameworks such as \texttt{Cobaya} \cite{Torrado2021}, which frequently employ nested samplers like \texttt{PolyChord} \cite{Handley2015} or standard Metropolis-Hastings algorithms to explore the cosmological parameter space. While these highly optimized pipelines work well for standard $\Lambda$CDM parameter estimation or global cosmological fits, they are mainly dedicated to late-time observables. Consequently, any arbitrary modifications to the thermal history in the MeV era, such as dynamical EDE components, goes beyond the scope of the aforementioned computational frameworks, and require a consistent restructuring of their coupled differential equations.

One of the important problems in Bayesian statistical analysis is that the  confidence intervals of the parameters strongly depend on the adopted priors shapes and ranges \cite{Ag, Gar, Mena}, a dependence that raises the question if statistical results can be reported in an unbiased and convincing way. Even though a prior-free assessment of confidence is not possible in general \cite{Ag}, approaches that try to circumvent this problem have been proposed and analyzed in the literature. One such possibility is based on the consideration of the $\mathcal{R}$ function \cite{Ag}, which factorizes the experimental evidence and the prior odds, and which can be interpreted geometrically as the shape distortion function of the probability density function. In \cite{Gar} it was shown that there is a simple way to obtain prior-independent constraints by using Bayesian analysis, and by using the $\mathcal{R}$ function. The function $\mathcal{R}$  is extremely useful when considering open likelihoods, that is, when data only constrain the value of the parameters from below or from above. The obtained formalism was applied  to the case of the analysis of  neutrino mass constraints from cosmology. Bayesian model comparison permits to consider the constraints coming from the different models, and to find  prior-independent and model-marginalized bounds \cite{Gar}. A novel method, which  provides robust limits that depend only on the considered dataset was developed in \cite{Mena}. It was shown that when considering several possible cosmological models, and by interpreting the Bayesian preference by using the Gaussian statistical evidence, the preferred model is least preferred as compared to   the two case model analysis. This approach was applied to the cosmological neutrino mass bounds, and for establishing the contribution of relic neutrinos to the dark matter density.

It is the main goal of the present investigation to consider  the implications of Early Dark Energy cosmological models on the Big Bang Nucleosynthesis processes, and to obtain robust observational constraints on the free parameters of the models.  In particular, we analyse four dark energy models, and we estimate their possible impact on the BBN processes. Firstly, we obtain an estimate of the maximum allowable value of the cosmological constant in the early Universe consistent with the nucleosynthesis data. Our results indicate that a much larger cosmological constant that presently observed may have existed in the early Universe, a result that strongly suggests that the cosmological constant may be a dynamical, time dependent quantity. Secondly, we investigate three distinct dark energy models, described with varying equations of state. The first model assumes a simple linear equation of state for the dark energy, while in a second model a polytropic equation of state is assumed. Finally, we consider a theoretical model in which dark energy is a temperature dependent quantity.

In order to obtain observational constraints on the Early Dark Energy models we adopt a direct approach by focusing primarily on the evolution of the primordial plasma and light element formation through \texttt{PRyMordial} \cite{PRyM, Burns2024}. Unlike standard BBN codes which treat beyond-Standard-Model physics as secondary perturbations, \texttt{PRyMordial} is designed to handle non-standard thermal histories, allowing for any modification to the Friedmann expansion and weak interaction rates. To explore the EDE parameter space and compare our theoretical predictions with recent astrophysical data, we couple our modified nucleosynthesis network to a nested sampling algorithm. We chose to use nested sampling not only because it can navigate non-Gaussian posterior distributions characteristic of dynamic dark energy models, but also because it computes the exact Bayesian evidence for the models considered. This allows for a consistent model comparison between the EDE frameworks considered in this analysis, namely the cosmological constant model and varying equation of state models, from linear to polytropic and temperature-dependent systems, and the standard BBN scenario.

	This paper is organized as follows. In Section~\ref{theory}, we present the Early Dark Energy scenarios considered in this study, together with the associated modifications to the first Friedmann equation in each framework. We then introduce the Big Bang Nucleosynthesis regime in Section~\ref{bbn}, where the computational approach is briefly discussed along with our strategy for estimating upper limits on model parameters through nucleosynthesis abundance ratios. Section~\ref{results} presents the results of our investigation, and we conclude in Section~\ref{final} with a statistical and physical discussion of the best-performing Early Dark Energy model.
	
	\section{Early Dark Energy models} \label{theory}
	
	In the following Section, we present the Early Dark Energy models considered in this work, emphasising on the evolution of the energy density and the specific equations of state for each candidate. We begin by defining the background evolution of a flat and homogeneous Friedmann-Lemaître-Robertson-Walker (FLRW)  universe and subsequently introduce the modifications to the Friedmann equations which arise from the inclusion of an early dark energy component. This framework provides the theoretical  basis for calculating the Hubble parameter, whose deviations will affect the temperatures relevant to the BBN epoch.
	
	\subsection{Standard Cosmology with a Cosmological Constant}
	In the standard framework, the homogeneous and isotropic Universe is described by the FLRW metric in flat Cartesian coordinates,
	\begin{equation}
		ds^2 = dt^2 - a(t)^2(dx^2 + dy^2 + dz^2),
	\end{equation}
	where $a(t)$ is the cosmic scale factor. in the following we adopt the natural system of units with $c=1$. The background dynamics are then governed by the Friedmann equations,
	\begin{equation}
		3H^{2}(t)=8\pi G\rho (t) +\Lambda,
	\end{equation}
	\begin{equation}
		2\dot{H}(t)+3H^{2(t)}=-8\pi G p(t)+\Lambda,
	\end{equation}
	where $H(t) = \dot{a}(t)/a(t)$ is the Hubble parameter defining the cosmic expansion rate, $G$ is the Newtonian gravitational constant, $\rho (t)$ is the total energy density, $p(t)$ is the total thermodynamic pressure, and $\Lambda$ is the cosmological constant.
	
	During the radiation-dominated era of the early Universe, the total energy density $\rho$ and pressure $p$ are dominated by relativistic species. These scale with the plasma temperature $T$ through the following relations
	\begin{equation}
		\rho (t)=\frac{4\sigma}{c} T^{4}(t), \quad p(t)=\frac{4\sigma}{3c} T^{4}(t),
	\end{equation}
	where  $\sigma$ denotes the Stefan-Boltzmann constant.
	The temporal evolution of these components is given by the standard energy conservation equation,
	\begin{equation}
		\dot{\rho}(t)+3H(t)\left[ \rho (t)+p(t)\right] =0.
	\end{equation}
	
	\subsection{Varying (Dynamical) Dark Energy}
	To explore deviations from the standard regime, we consider models of dynamical dark energy, in which the dark energy density is not a constant but is a function of time. The corresponding modified Friedmann equations introduce an effective total density $\rho_{eff}$ and effective total pressure $p_{eff}$, which include the dynamic dark energy components $\rho_{DE}(t)$ and $p_{DE}(t)$,
	\begin{equation}
		3H^{2}(t)=8\pi G\left[ \rho (t) +\rho_{DE}(t)\right] =8\pi G\rho _{eff}(t),
	\end{equation}
	\begin{equation}
		2\dot{H}(t)+3H^{2}(t)=-8\pi G\left[ p(t)+p_{DE}(t)\right] =-\frac{8\pi G}{c^{2}}p_{eff}(t).
	\end{equation}
	Assuming the standard relations for the radiation bath hold, the global energy conservation equation becomes
	\begin{equation}
		\dot{\rho}_{eff}(t)+3H(t)\left[ \rho _{eff}(t)+p_{eff}(t)\right] =0.
	\end{equation}
	Expanding this relation results in the following global conservation equation
	\bea
		\dot{\rho}(t)&+&3H(t)\left[ \rho (t)+p(t)\right] \nonumber\\
&+&\dot{\rho}_{DE}(t)+3H(t)\left[ \rho _{DE}(t)+p_{DE}(t)\right] =0.
	\eea
	
For the purposes of this study, we assume a non-interacting model in which the dark energy component is minimally coupled to the standard plasma, meaning there is no direct energy-momentum exchange between $\rho_{DE}$ and the radiation bath. The two components influence each other only through their cumulative contribution to the gravitational background, meaning through the Hubble expansion rate. Consequently, the energy of the radiation and dark energy components are independently conserved
	\begin{equation}
		\dot{\rho}(t)+3H(t)\left[ \rho(t) +p(t)\right] =0,
	\end{equation}
	\begin{equation}
		\dot{\rho}_{DE}(t)+3H(t)\left[ \rho _{DE}(t)+p_{DE}(t)\right] =0.
	\end{equation}
	
	The temporal evolution of the dark energy density is entirely determined by its assumed equation of state. We consider three distinct theoretical cases to model this dynamical behaviour.
	
	\subsubsection{Linear Equation of State}
	The simplest Early Dark Energy extension of the standard $\Lambda$CDM model is obtained by assuming a linear equation of state for dark energy, where the pressure of the dark energy component is strictly proportional to its energy density through a constant equation of state parameter $w$
	\begin{equation}
		p_{DE}(t)=w \rho _{DE}(t),
	\end{equation}
	where we specifically restrict $w \in (-1, 0)$ to explore the quintessence regimes. In quintessence models, a slowly rolling scalar field acts as the dynamical dark energy, which requires a negative pressure that departs from a constant cosmological constant scenario with $w = -1$. Substituting the equation of state into the dark energy conservation equation gives
	\begin{equation}
		\dot{\rho}_{DE}(t)+3(1+w)H(t)\rho _{DE}(t)=0,
	\end{equation}
	which integrates to provide the scaling of the dark energy density with the scale factor of the Universe as
	\begin{equation}
		\rho _{DE}(t)=\rho _{DE,0}a^{-3\left( 1+w\right) }(t).
	\end{equation}
	
Since  we do not directly relate this early dark energy fluid to late-time cosmological constant values, $\rho _{DE,0}$ is considered an unknown density scale parameter of the model. Next, we rewrite the radiation energy density component as a function of the present-day Cosmic Microwave Background temperature $T_0$,
	\begin{equation}
		\rho_{0} = \frac{4\sigma}{c} T_{0}^{4},
	\end{equation}
	so that we rewrite the radiation energy density component as a function of the scale factor $a(t)$ as
	\begin{equation}
		\rho (t)=\frac{\rho _{0}}{a^{4}(t)}, \quad T(t)=\frac{T_{0}}{a(t)}.
	\end{equation}
	The modified cosmological evolution equation becomes,
	\begin{equation}
		3H^{2}(t)=\frac{8\pi G\rho _{0}}{a^{4}(t)}+\frac{8\pi G\rho_{DE,0}}{a^{3(1+w)}(t)}.
	\end{equation}
	
	\subsubsection{Polytropic Equation of State}
	
Alternatively, we consider a polytropic equation of state for the Early Dark Energy, a formulation frequently used to model various physical fluids. The pressure depends non-linearly on the energy density, governed by a polytropic constant $K$ and an index $\gamma$,
	\begin{equation}
		p_{DE}(t) = K\rho_{DE}^{\gamma}(t).
	\end{equation}
	Under this assumption the conservation equation for dark energy leads to the following relation
	\begin{equation}
		\dot{\rho}_{DE}(t) + 3H(t)\left[\rho_{DE}(t) + K\rho_{DE}^{\gamma}(t)\right] = 0,
	\end{equation}
	from which one can easily obtain
	\begin{equation}
		\int \frac{d\rho_{DE}}{\rho_{DE} + K\rho_{DE}^{\gamma}} = -3\ln a.
	\end{equation}
	
Substituting $u = \rho_{DE}^{1-\gamma}$, the left-hand side evaluates to $\frac{1}{1-\gamma}\ln(\rho_{DE}^{1-\gamma}+K)$, giving
	\begin{equation}
		\rho_{DE}^{1-\gamma} + K = \tilde{C}\, a^{3(\gamma-1)},
	\end{equation}
	where $\tilde{C}$ is an integration constant. Setting $C = 1/\tilde{C}$ and rearranging, the early dark energy density profile becomes
	\begin{equation}
		\rho_{DE}(t) = \left[\frac{a^{3(\gamma-1)}(t)}{C} - K\right]^{1/(1-\gamma)}.
	\end{equation}
	The corresponding Hubble evolution equation is modified as
	\begin{equation}
		3H^2(t) = \frac{8\pi G\rho_{0}}{a^{4}(t)} + 8\pi G\left[\frac{a^{3(\gamma-1)}(t)}{C} - K\right]^{1/(1-\gamma)}.
	\end{equation}
	
	Rather than treating the polytropic index $\gamma$ as a continuous free parameter, we fix it to discrete values, so that we restrict our analysis to the following cases:
	\begin{itemize}
		\item $\gamma = 4/3$ -- we consider that the dark energy component has a radiation-like equation of state, but which doesn't interact with the radiative sector during the BBN era
		\item $\gamma = 2$ -- a stiff equation of state causes the dark energy density to dilute faster than the background radiation, decaying rapidly to avoid late-time cosmological constraints violations.
	\end{itemize}
	
	\subsubsection{Temperature-Dependent Equation of State}
	Finally, we explore a scenario where the dark energy equation of state evolves with temperature $T$ alongside the thermal bath of particles, 
	\begin{equation}
		\rho_{DE}(T) = \rho_{DE,0} \left( \frac{T}{T_0} \right)^{3(1+w(T))},
	\end{equation}
	with $T_0 = 2.7255$ K being the present-day CMB temperature.
	The equation of state is parameterized as follows
	\begin{equation}
		p_{DE}=w(T)\rho _{DE},
	\end{equation}
	for which various functional forms can be assumed. We impose a linear dependence on temperature, given by a coupling constant $\alpha$, that can be expressed as
	\begin{equation}
		w(T)=w_{0}+\alpha T=w_{0}+\frac{\alpha T_{0}}{a}, 
	\end{equation}
	where we take $w_{0}=-1$ so that in the limit $\alpha \rightarrow 0$ we recover the standard cosmological constant regime.
	
	\section{Nucleosynthesis Modeling} \label{bbn}
	
	We revise in the following Section the Big Bang Nucleosynthesis process together with the numerical procedure used for simulating the formation of abundance ratios following the modifications of the freeze-out temperature. We describe the integration of our EDE framework into the standard BBN reaction network and we discuss the strategy for numerical estimations of upper bounds. We perform a Bayesian inference through a nested sampling technique, which allows us to map the multidimensional posterior distributions of our model parameters against observed primordial abundances.
	
	\subsection{Big Bang Nucleosynthesis Theory}
	Big Bang Nucleosynthesis describes the production of the lightest elements such as Hydrogen, Deuterium, Helium-4, and Lithium-7 during the first few minutes of the Universe, which are highly sensitive to the expansion rate. In the presence of a dark energy component, the Hubble rate $H$ expressed from the first Friedmann equation is modified to include the effective energy density,
	\begin{equation}
		H(T) = \sqrt{\frac{8\pi G}{3c^2} \rho_{eff}},
	\end{equation}
	which evolves in dynamic dark energy models.
	The expansion rate competes with the weak interaction rates $\Gamma_{n \leftrightarrow p}$, which maintain chemical equilibrium between neutrons and protons through the nuclear processes
	\begin{equation}
		n + \nu_e \rightarrow p + e^-, \quad n + e^+ \rightarrow p + \bar{\nu}_e, \quad n \rightarrow p + e^- + \bar{\nu}_e.
	\end{equation}
	The weak interaction rate scales approximately as $\Gamma_{n \leftrightarrow p} \propto G_F^2 T^5$ and it decreases as the Universe cools, reaching the so-called freeze-out regime in which the expansion rate $H(T)$ exceeds the interaction rate $\Gamma$,
	\begin{equation}
		\Gamma_{n \leftrightarrow p}(T_f) \approx H(T_f),
	\end{equation}
	where $T_f$ represents the freeze-out temperature.
	Due to the presence of the EDE component, $H(T)$ can increase which determines the freeze-out to occur earlier, corresponding to a higher $T_f$. The neutron-to-proton ratio at freeze-out is governed by the Boltzmann factor
	\begin{equation}
		\left( \frac{n}{p} \right)_{f} = \exp\left( -\frac{(m_n - m_p) c^2}{k_B T_f} \right),
	\end{equation}
	such that a higher $T_f$ results in a larger $(n/p)_f$ ratio, which subsequently increases the final primordial helium mass fraction, approximately given by
	\begin{equation}
		Y_p \approx \frac{2(n/p)}{1 + (n/p)}.
	\end{equation}
	
	\begin{table}[h!]
		\centering
		\begin{tabular}{|l|c|c|c|}
			\hline
			\text{Isotope} & \text{Observable} & \text{Observed Value} & \text{SBBN Prediction} \\ \hline \hline
			Helium-4 & $Y_p$ & $0.245 \pm 0.003$ & $0.247 \pm 0.0001$ \\ \hline
			Deuterium & D/H $\times 10^5$ & $2.527 \pm 0.030$ & $2.51 \pm 0.11$ \\ \hline
			Helium-3 & $^3$He/H $\times 10^5$ & $1.1 \pm 0.2$ & $1.0 \pm 0.1$ \\ \hline
			Lithium-7 & $^7$Li/H $\times 10^{10}$ & $1.6 \pm 0.3$ & $4.7 \pm 0.7$ \\ \hline
		\end{tabular}
		\caption{Primordial abundances from PDG and Standard BBN predictions.}
		\label{abundances}
	\end{table}
	
	This detection of the primordial abundances with high sensitivity in databases such as the Particle Data Group \cite{pdg} (see Table~\ref{abundances}) allows us to determine strict limits on the EDE density.
	In this analysis, the parameter estimation process will rely exclusively on the primordial helium mass fraction $Y_p$, deuterium and $^3$He abundance. We exclude Lithium-7 from the $\chi^2$ likelihood calculations as the Lithium problem is highly debated in the literature, since Standard BBN theory currently predicts a $^7$Li abundance approximately three times higher than what is observed in metal-poor halo stars. 
	
	To simulate the synthesis of nuclei in the early Universe, we develop a program onto the \texttt{PRyMordial} package \cite{PRyM, Burns2024}, which integrates the Boltzmann equations for primordial nuclear species from a high temperature of 10 MeV to 0.001 keV. The code simultaneously evaluates the evolution of the cosmic scale factor, baryon number density, and the distinct thermal baths of photons and neutrinos, solving at each step a stiff system of coupled ordinary differential equations describing both the rapid thermonuclear reaction networks and the macroscopic cooling of the expanding Universe. This cosmological state vector describes efficiently the Standard Model phenomena, including incomplete neutrino decoupling and finite-temperature quantum electrodynamic effects, which result in  accurate abundance predictions.
	
	The advantage of using  \texttt{PRyMordial} is that it contains a \texttt{New Physics} interface that can accommodate diverse modifications in the background equations and plasma sector, coming from various theoretical scenarios. Rather than requiring beyond-Standard-Model extensions to be defined in terms of predefined parameters, the code allows for a direct modification of the thermodynamic quantities driving the integration. By an explicit deviation of the Hubble expansion rate or injecting dynamic energy densities into the plasma, the software computes from the initial nuclear rates the way in which these high-energy deformations shift the neutron freeze-out temperature. It then outputs theoretical predictions for observable light element results, such as the helium mass fraction deuterium and Helium-3 abundance ratios, which are the required quantities that allow for direct statistical comparison against astrophysical data.
	
	\subsection{PRyMordial Framework Integration}
	We embed our modified Friedmann equations directly into the \texttt{PRyMordial} differential equation solver, treating the EDE contribution as an additional energy density component in the early Universe background.
	
	The radiation energy density at the present epoch, $\rho_0$, is a known quantity determined from the Cosmic Microwave Background measurements of temperature $T_0 \approx 2.725$ K. In contrast, the EDE component $\rho_{DE,0}$ in the linear case or the constant $C$ in the polytropic case cannot be directly inferred by observational data, which is due to the non-interacting nature of the models considered. As the dark energy is minimally coupled to the standard plasma, it evolves independently according to its own equation of state and because this EDE is a transient phenomenon, its amplitude is not fixed by late-time $\Lambda$ constraints and must be searched for as a free parameter.
	
	Consequently, for the cosmological constant model we search for upper bounds for the early $\Lambda$ parameter, while in the case of the linear time-varying model, the density scale $\rho_{DE,0}$ and the equation of state parameter $w$ are searched for. In the polytropic model, the model parameters are estimated when $\gamma$ is fixed to two distinct physical cases. Finally, in the temperature-dependent model, the $\alpha$ parameters is also constrained by the BBN  primordial abundance data.
	
	We integrate through our model interface to \texttt{PRyM}'s ODE system an additional dynamic scale factor calculation from an extended background integration. The cumulative temperature mapping as a function of cosmic time can introduce the dark sector energy density described as a function of time into the total energy density calculation in \texttt{PRyMordial}, which is natively temperature-based. This functionality is needed as the reparametrization of $\rho_{EDE}$ in terms of time requires a recalculation of the scale factor $a(t)$, which becomes a dynamical variable in the Friedmann equation. In practice, the cosmological constant and temperature-dependent equation of state will not need this recomputation of the scale factor, but the linear and polytropic models will, as a consequence of altering the background dynamics through an explicit $\rho_{EDE}(t)$ evolution.
	
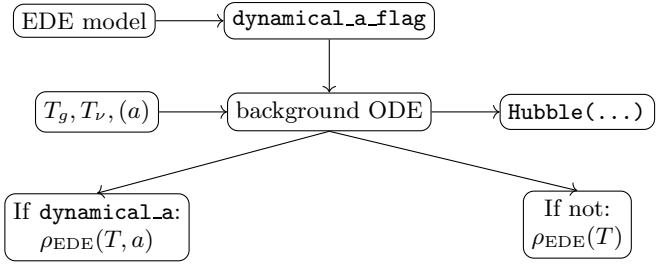
\begin{figure}[htbp!]
	\begin{tikzpicture}[font=\small, node distance=4mm, every node/.style={align=center}]
		\node (eden) [draw, rounded corners, minimum width=8mm] {EDE model};
		\node (flag) [draw, rounded corners, right=9mm of eden, minimum width=9mm] {\texttt{dynamical\_a\_flag}};
		\node (ode) [draw, rounded corners, below=7mm of flag, minimum width=11mm] {background ODE};
		\node (state) [draw, rounded corners, left=9mm of ode, minimum width=9mm] {$T_g, T_\nu, (a)$};
		\node (hub) [draw, rounded corners, right=9mm of ode, minimum width=10mm] {\texttt{Hubble(...)}};
		\draw[->] (eden) -- (flag);
		\draw[->] (flag) -- (ode);
		\draw[->] (state) -- (ode);
		\draw[->] (ode) -- (hub);
		\node (on) [draw, rounded corners, below=8mm of state, minimum width=12mm] {If \texttt{dynamical\_a}:\\ $\rho_{\rm EDE}(T,a)$};
		\node (off) [draw, rounded corners, below=8mm of hub, minimum width=12mm] {If not:\\ $\rho_{\rm EDE}(T)$};
		\draw[->] (ode.south) -- (on.north);
		\draw[->] (ode.south) -- (off.north);
	\end{tikzpicture}
	\caption{Integration scheme for the EDE models into the \texttt{PRyMoridal} package.}
	\label{scheme}
	\end{figure}

	Within \texttt{PRyM/PRyM\_main.py}, the background solver defines the right-hand side of the ODE system \texttt{dTtotdt}, and evaluates the \texttt{PRyMini.dynamical\_a\_flag}, passed as a model attribute for each EDE instance. When enabled, the state vector contains both the scale factor, and the corresponding evolution equation, $da/dt = a \times \texttt{Hubble(...)}$, so that the Friedmann term can be evaluated using this dynamic scale factor. Consequently, the \texttt{Hubble} function computes the total energy density and incorporates the EDE contribution through the \texttt{PRyMthermo.rho\_EDE(Tg, a)} function. Conversely, if the dynamical flag is disabled, the scale factor is not evolved and $\rho_{\text{EDE}}$ is inherently treated as independent of $a$. The complete implementation of this modified framework is publicly available in the \texttt{eden} repository (\url{https://github.com/croi900/eden}), and the implementation scheme can be visualized in Fig.~\ref{scheme}.
	
	The parameter estimation is performed by comparing the theoretical mass fractions generated by the \texttt{PRyMordial} ODE solver against the observed primordial abundances. This comparison is handled by a log-likelihood function derived from a $\chi^2$-test constructed by summing the squared residuals of the light element abundances $\theta$, weighted by their respective observational uncertainties $\sigma$,
	\begin{equation}
		\chi^2(\theta) = \sum_{i} \frac{(X_{i, pred}(\theta) - X_{i, obs})^2}{\sigma_i^2}
	\end{equation}
	where $X_i \in \{Y_p, D/H, ^3He/H\}$.
	The final log-likelihood is expressed as
	\begin{equation}
		\ln \mathcal{L} = -0.5 \; \chi^2(\theta).
	\end{equation}
	The resulting samples including the physical parameters, predicted abundances and the corresponding log-likelihood are recorded iteratively to map the posterior distribution.
	
	\subsection{Nested Sampling Algorithm}
		
	The statistical challenge in constraining early dark energy parameters arises because the theoretical abundances converge to SBBN values as the dark energy contribution vanishes, creating a broad $\chi^{2}$ plateau in the likelihood function. Because the simulated results are highly consistent with observational data, standard Markov Chain Monte Carlo (MCMC) samplers remain trapped in this plateau, searching the lower prior volume without identifying a significant upper bound. To efficiently evaluate the posterior distribution for precise parameter limits, a nested sampling approach is further applied, as it can efficiently search over such plateaus.
	
	In Bayesian processes, we draw inferences regarding a set of parameters $\theta$ for a given model $M$ and observed data $D$ by relying on Bayes' theorem,
	\begin{equation}
		\mathcal{P}(\theta | D, M) = \frac{\mathcal{L}(D | \theta, M)\; \pi(\theta | M)}{\mathcal{Z}(D | M)}
	\end{equation}
	where $\mathcal{P}(\theta | D, M)$ is the posterior probability distribution, $\mathcal{L}(D | \theta, M)$ is the likelihood function, and $\pi(\theta | M)$ is the prior probability density. 
	
	The denominator, $\mathcal{Z}(D | M)$, is called the Bayesian evidence or marginal likelihood, serving as a normalization constant for the posterior. This quantity is defined as the integral of the likelihood over the entire prior parameter space,
	\begin{equation}
		\mathcal{Z} = \int \mathcal{L}(\theta) \pi(\theta) d\theta.
	\end{equation}
	While the evidence is unnecessary for simple parameter estimation drawn from standard MCMC methods, it is the fundamental quantity required for Bayesian model selection, as two competing models can be compared through the Bayes factor, $K = \frac{\mathcal{Z}_1}{\mathcal{Z}_2}$.
	
	When dealing with multi-dimensional systems, the evidence integral becomes increasingly hard to estimate using standard grid or Monte Carlo integration. Hence, the nested sampling technique, introduced in \cite{Skilling, Ashton}, resolves to this issue by mapping the $D$-dimensional parameter space onto a one-dimensional domain defined by the prior volume $X(\lambda)$, which is the prior mass $X$ with likelihoods strictly greater than a threshold $\lambda$,
	\begin{equation}
		X(\lambda) = \int_{\mathcal{L}(\theta) > \lambda} \pi(\theta) d\theta,
	\end{equation}
	ranging from $1$ denoting the entire prior volume to $0$, where the likelihood is maximal. The evidence integral is therefore rewritten as a one-dimensional integral over the prior volume,
	\begin{equation}
		\mathcal{Z} = \int_{0}^{1} \mathcal{L}(X) dX,
	\end{equation}
	evaluated through an algorithm which maintains a collection of $N$ so-called live points drawn from the prior distribution $\pi(\theta)$: 
	\begin{enumerate}
		\item At iteration $i$, identify the live point $\theta_i$ with the lowest likelihood, denoted as $\mathcal{L}_i$.
		\item Remove this point from the active set and record its coordinates and likelihood.
		\item Estimate the shrinkage of the prior volume; on average, the maximum of $N$ random variables drawn uniformly from $[0, X_{i-1}]$ shrinks the volume by a factor of $\exp(-1/N)$, so that $X_i \approx \exp(-i/N)$.
		\item Compute the weight of the removed point, $w_i = X_{i-1} - X_{i}$, and add its contribution to the numerical approximation of the evidence  $\mathcal{Z} \approx \sum \mathcal{L}_i w_i$.
		\item Sample a new point from the prior distribution, imposing the strict constraint that its likelihood must exceed $\mathcal{L}_i$. 
	\end{enumerate}
	
	The iteration continues until a predefined convergence criterion is met, when the remaining evidence is a negligibly small fraction of the accumulated evidence $\mathcal{Z}$. 
	As the sequence of prior volumes $X_i$ is probabilistic, nested sampling contains a statistical uncertainty given as the log-evidence, $\ln(\mathcal{Z})$. The previously presented algorithm is implemented efficiently in the \texttt{dynesty} package \cite{dynesty}, which allows us to estimate the model parameter's upper bounds.
	
	\section{Implementation and Results} \label{results}
	
	The computational strategy consists of integrating the Early Dark Energy models into the \texttt{PRyMordial} nucleosynthesis solver with a nested sampling framework to map the early dark energy parameter space, implemented in \texttt{eden} framework. To perform the multidimensional parameter estimation and calculate the Bayesian evidence, we use the \texttt{dynesty} nested sampling package and for the subsequent posterior analysis and visualizations we use the \texttt{GetDist} library. In this section, we summarize the prior distributions  and the resulting statistical constraints for EDE models.

	\subsection{The Cosmological Constant Model}
	
	To ensure the BBN network is computed within physical and observational boundaries, we impose informative Gaussian priors on four parameters, which are included as external constraints from CMB data. The standard plasma sector is treated identically across all EDE models and includes the neutron lifetime $\tau_n = 879.4 \pm 0.6$ s, the baryon density $\Omega_{\mathrm{b}}h^2 = 0.02230 \pm 0.00015$, and two nuclear reaction rate uncertainty parameters $p_{\mathrm{np\gamma}} = 0 \pm 1$ and $p_{\mathrm{dp}{}^3\mathrm{He}\gamma} = 0 \pm 1$, which randomize the helium and deuterium abundance rates. The latter two are dimensionless parameters that shift the thermonuclear reaction rates $n + p \to d + \gamma$ and $d + p \to {}^3\mathrm{He} + \gamma$ within their experimental uncertainty regions from the NACRE II database, where $p = 0$ corresponds to the median rate and $p = \pm 1$ to a one-sigma deviation \cite{nacre2}.
	
	For the \texttt{CC} model, the dark energy sector contains a single parameter $\Lambda$ to which we apply a log-uniform prior to explore its influence over multiple orders of magnitude,
	\begin{equation}
		\log_{10}\Lambda \sim \mathcal{U}(-20,\,-4).
	\end{equation}
	
	We initialized our prior interval over the above scales with the aim to obtain a maximally allowed value of the cosmological constant through the evidence $\mathcal{Z}$ estimation, and not a Gaussian posterior. 
	Following the nested sampling evaluation, the Bayesian evidence for the \texttt{CC} model was calculated as $\ln \mathcal{Z}_{\mathrm{CC}} = -1.2905 \pm 0.0441$. The evidence uncertainty indicates that there is less than a 5\% probability of obtaining a value for the logarithm of the evidence that deviates significantly from this estimate. The marginalized posterior constraints for the dark energy parameter include the 68\% and 95\% upper limit, summarized in Table \ref{cc_posteriors}. The prior ranges for the current and following models are chosen based on the computational limits of \texttt{PRyMordial}. All credible intervals and upper limits are conditioned by the prior ranges and forms specified for each EDE model, influencing the posterior under those assumptions, which are not frequentist confidence limits independent of the prior support.
	
	\begin{table}[h!]
		\centering
		\renewcommand{\arraystretch}{1.3}
		\begin{tabular}{lcc}
			\hline
			\text{Parameter} & \text{68\% CI} & \text{95\% CI} \\
			\hline
			$\Lambda$ $[\mathrm{MeV}^4]$ &  $[4.24 \times 10^{-19},\, 9.30 \times 10^{-13}]$ & $9.41 \times 10^{-12}$ \\
			\hline
		\end{tabular}
		\caption{Marginalized posterior constraints for the CC model.}
		\label{cc_posteriors}
	\end{table}
	
	While the 68\% credible interval provides a measure of the posterior's central mass, the 95\% upper limit is the preferred metric for setting hard constraints on the parameter space. For the \texttt{CC} model, the constraint $\Lambda < 9.41 \times 10^{-12}~\text{MeV}^4$ represents a physical geometric limit of $\Lambda < 4.07 \times 10^{-33}~\text{cm}^{-2}$.
	
\begin{figure}[htbp!]
		\centering
		\includegraphics[scale = 0.52]{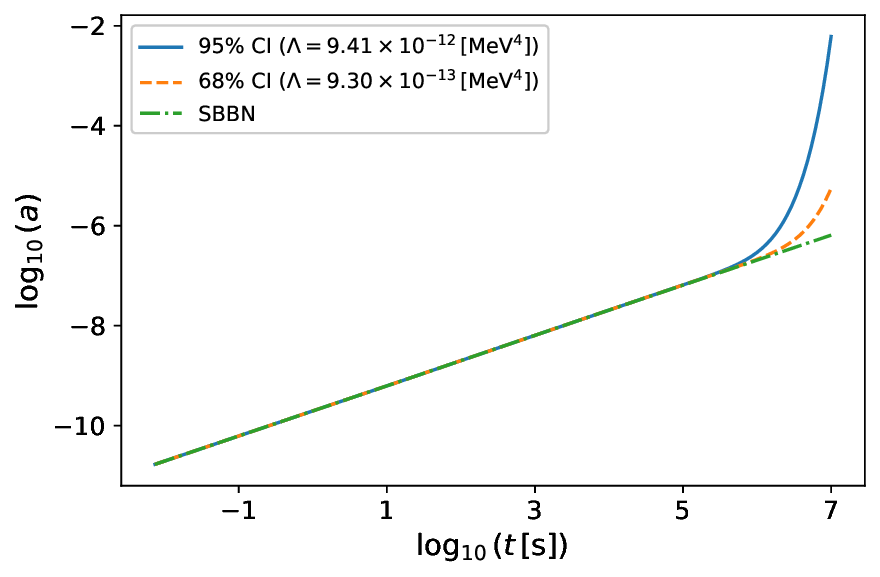}
		\includegraphics[scale = 0.52]{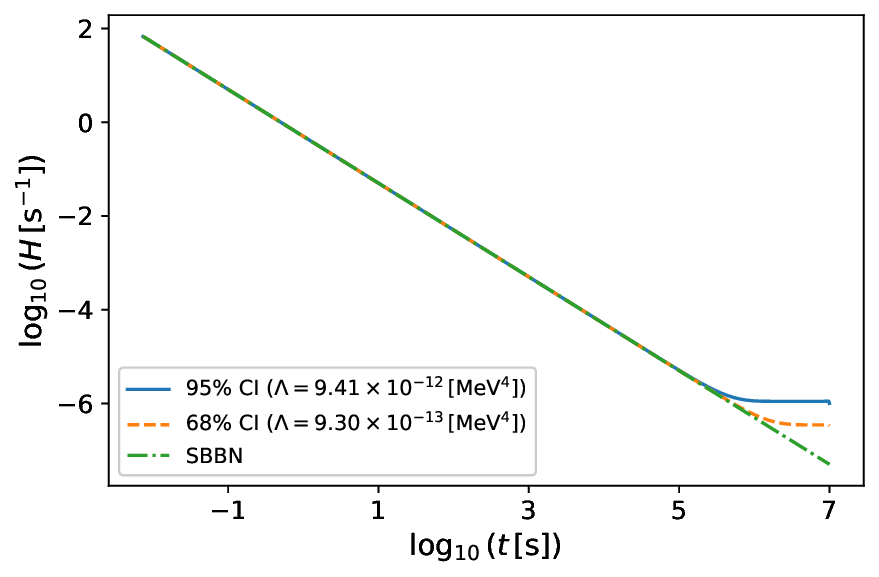}
		\caption{Evolution of the logarithm of the scale factor and the Hubble rate affected by the $\Lambda$ parameter estimated at an upper limit of 68\% and 95\% CL, compared to the SBBN predictions.}
		\label{aoft_CC}
	\end{figure}

	The effects of our upper bound on the cosmological constant $\Lambda$ are shown in Fig.~\ref{aoft_CC}, which tracks the logarithmic evolution of both the scale factor (top panel) and the Hubble rate (bottom panel) against the standard simulation trend obtained with no \texttt{New Physics} contribution in PRyMordial. At the 95\% confidence limit, the scale factor deviates from standard radiation-dominated growth, increasing exponentially faster then SBBN as the cosmological constant component starts to dominate. This behaviour is confirmed in the second plot, where the Hubble rate reaches a plateau before the standard limit is achieved, implying that the Universe enters too early a vacuum-dominated phase. This behaviour implies that if $\Lambda$ were any larger than this limit, the weak interactions will freeze out too early, leading to much higher neutron-to-proton ratio and therefore absurdly large nuclei abundances.
	
	\begin{figure}[htbp!]
		\centering
		\includegraphics[scale=0.22]{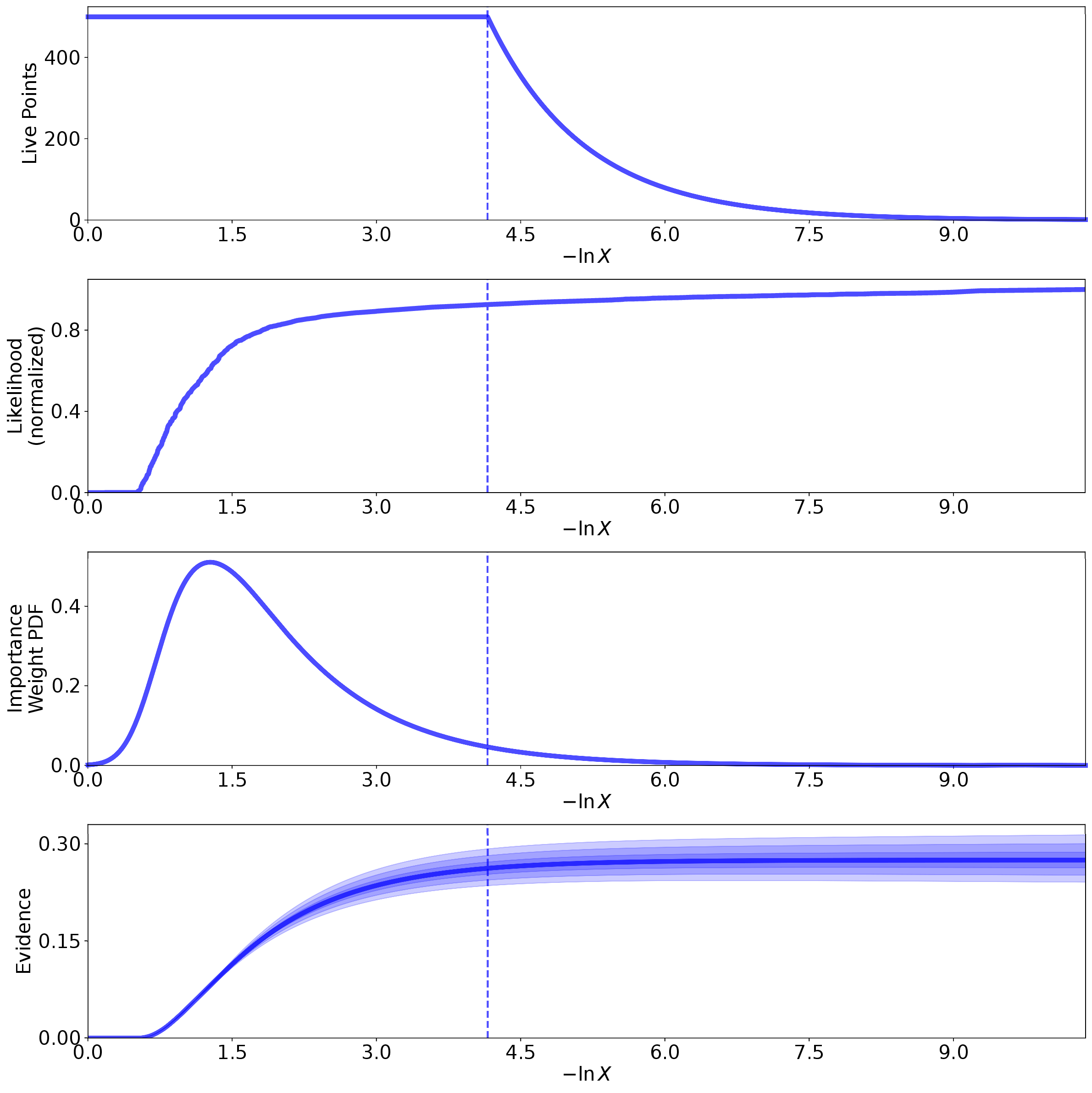}
		\caption{Statistical summary of the nested sampling for the \texttt{CC} model.}
		\label{stat_CC}
	\end{figure}
	
	We can verify that the obtained limits are numerically stable by computing the nested sampling diagnostics, shown in Fig.\ref{stat_CC}. As the algorithm compresses the prior volume from left to right though $-\ln X$, the number of live points remains constant initially and then decays as the sampler narrows the parameter search space. The normalized likelihood increases fast up to a flat plateau, representing the parameter space that approaches the standard limit. The real proof of convergence is the importance weight curve, which forms a single peak at the beginning of the search and flattens at the end, demonstrating that the Bayesian evidence was successfully localized.
	
\begin{figure}[htbp!]
		\centering
		\includegraphics[scale =0.5]{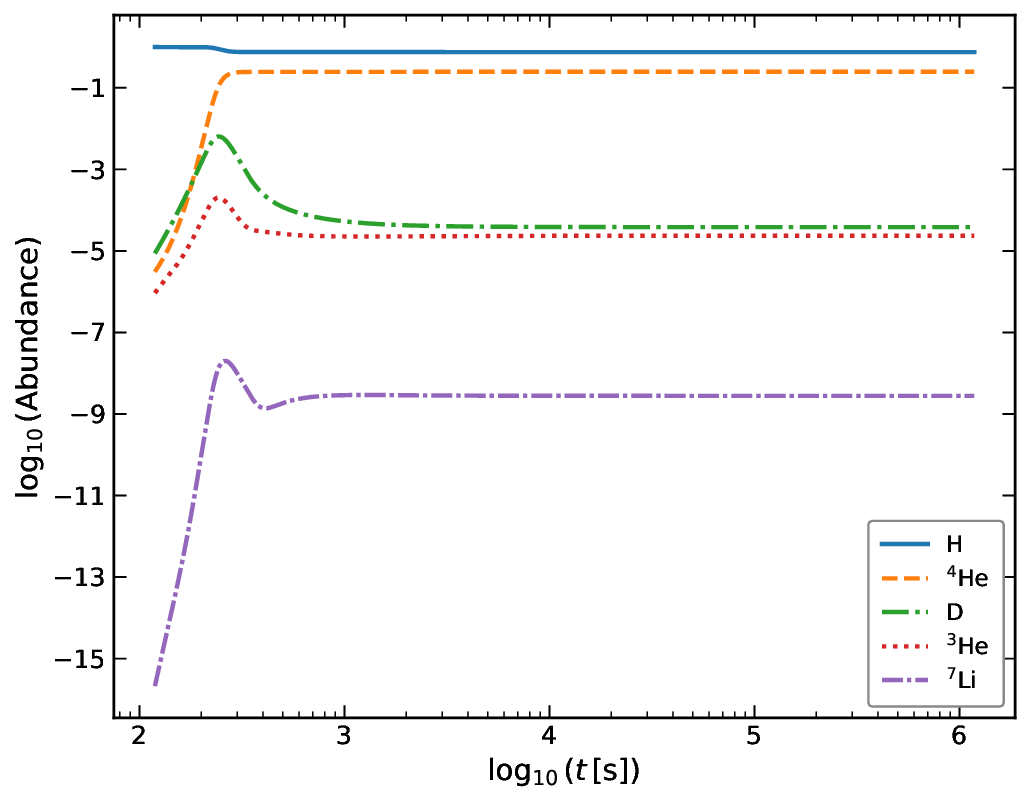}
		\caption{The log-log time evolution of light elements  for the \texttt{CC} model.}
		\label{Yi_CC}
	\end{figure}

	To quantify the time dependence of light-element production, we show in Fig.~\ref{Yi_CC} the log-log evolution of all light abundances for the \texttt{CC} model at the 95\% upper-limit parameters. While $^4$He reaches a plateau quickly due to its high binding energy relative to the other elements, D, $^3$He, and $^7$Li display a peak near $\log_{10} t \simeq 2.4$ ($t \sim 10^2$--$10^3\,\mathrm{s}$) followed by a decay because of a high reactivity, stabilizing at their relic values. This behaviour is consistent with the standard abundance evolution, as no strong distortion exists in the nucleosynthesis process.

	\begin{figure}[htbp!]
		\centering
		\includegraphics[scale =0.23]{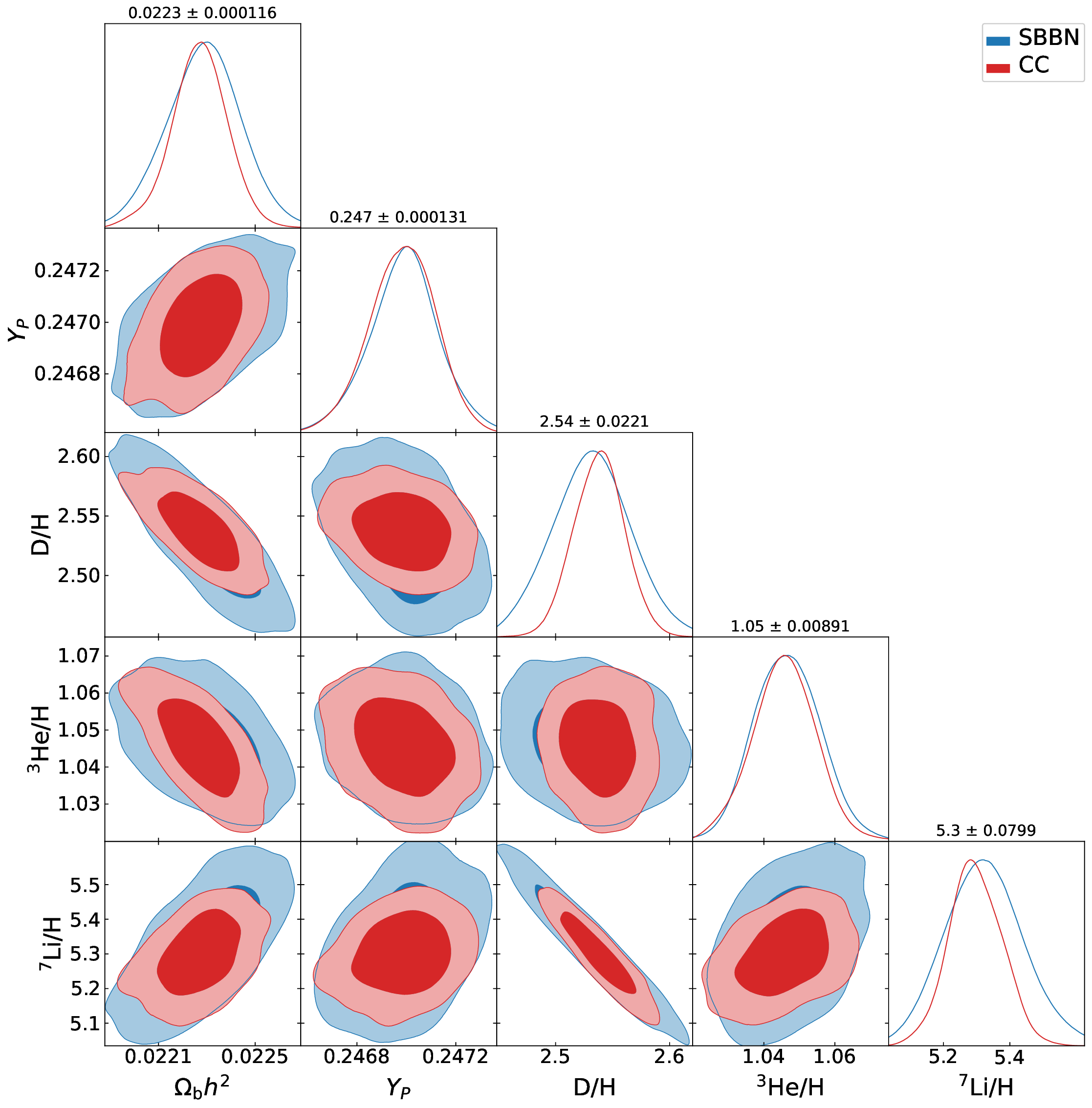}
		\caption{Nuclear abundances together with $\Omega_b h^2$ for the \texttt{CC} model overlaid with SBBN results from PRyMordial.}
		\label{abund_CC}
	\end{figure}
	
	Finally, the abundance corner plots in Fig.\ref{abund_CC} show nearly identical overlap between our constrained early dark energy models and the standard BBN bounds. This behaviour is expected as we searched for an upper limit rather than claiming a bounded estimation.

	\subsection{The Linear EDE Model}
	
	The \texttt{Linear} model introduces a dynamic fluid characterized by two free parameters, namely the dark energy density $\rho_{DE,0}$ and a constant equation of state parameter $w$. We restrict $w$ to the quintessence regime and apply log-uniform and uniform priors, respectively
	\begin{align}
		\log_{10}\rho_{DE,0} &\sim \mathcal{U}(-30,\,-2), \\
		w &\sim \mathcal{U}(-1,\,0).
	\end{align}
	
	The nested sampling evaluation calculated a Bayesian evidence of $\ln \mathcal{Z}_{\mathrm{Linear}} = -2.0065 \pm 0.0593$, with the resulting posterior constraints for the dynamic dark energy parameters detailed in Table \ref{linear_posteriors}.
	
	\begin{table}[h!]
		\centering
		\renewcommand{\arraystretch}{1.3}
		\begin{tabular}{lcc}
			\hline
			\text{Parameter}  & \text{68\% CI} & \text{95\% CI} \\
			\hline
			$\rho_{DE,0}$ $[\mathrm{MeV}^4]$ & $[9.42 \times 10^{-29},\, 4.36 \times 10^{-17}]$ & $2.45 \times 10^{-13}$ \\
			$w$ &  $[-0.918,\, -0.432]$ & $-0.269$ \\
			\hline
		\end{tabular}
		\caption{Marginalized posterior constraints for the Linear  equation of state EDE model.}
		\label{linear_posteriors}
	\end{table}
	
	For the \texttt{Linear} model, the 95\% upper limit $\rho_{DE,0} < 1.06 \times 10^{-34}~\text{cm}^{-2}$ represents the maximum possible energy density of the dark sector that could have contributed during the BBN epoch without perturbing the primordial abundances beyond their observational uncertainties. As the energy density scales as $a^{-3(1+w)}$, an equation of state parameter closer to zero would cause the dark sector's density to act as pressureless matter.
	
	\begin{figure}[htbp!]
		\centering
		\includegraphics[scale = 0.52]{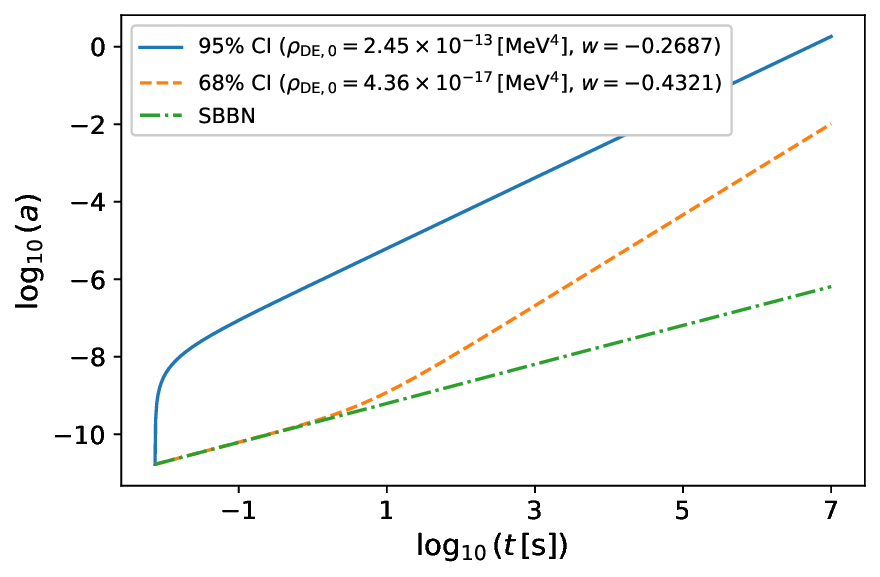}
		\includegraphics[scale = 0.52]{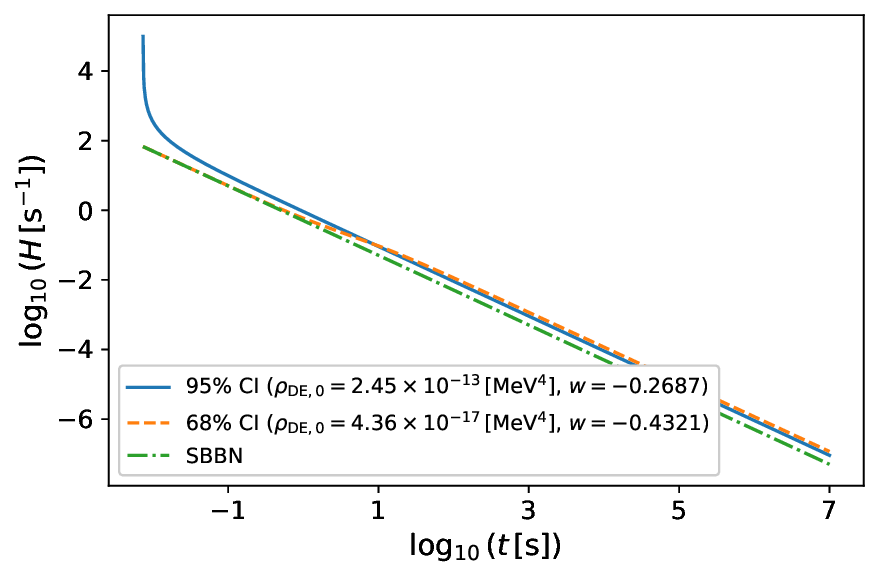}
		\caption{Evolution of the logarithm of the scale factor and the Hubble rate affected by the \texttt{Linear} equation of state model with parameters $\rho_{DE,0}$ and $w$ compared to the SBBN trends.}
		\label{aoft_linear}
	\end{figure}
	
	\begin{figure}[htbp!]
		\centering
		\includegraphics[scale=0.22]{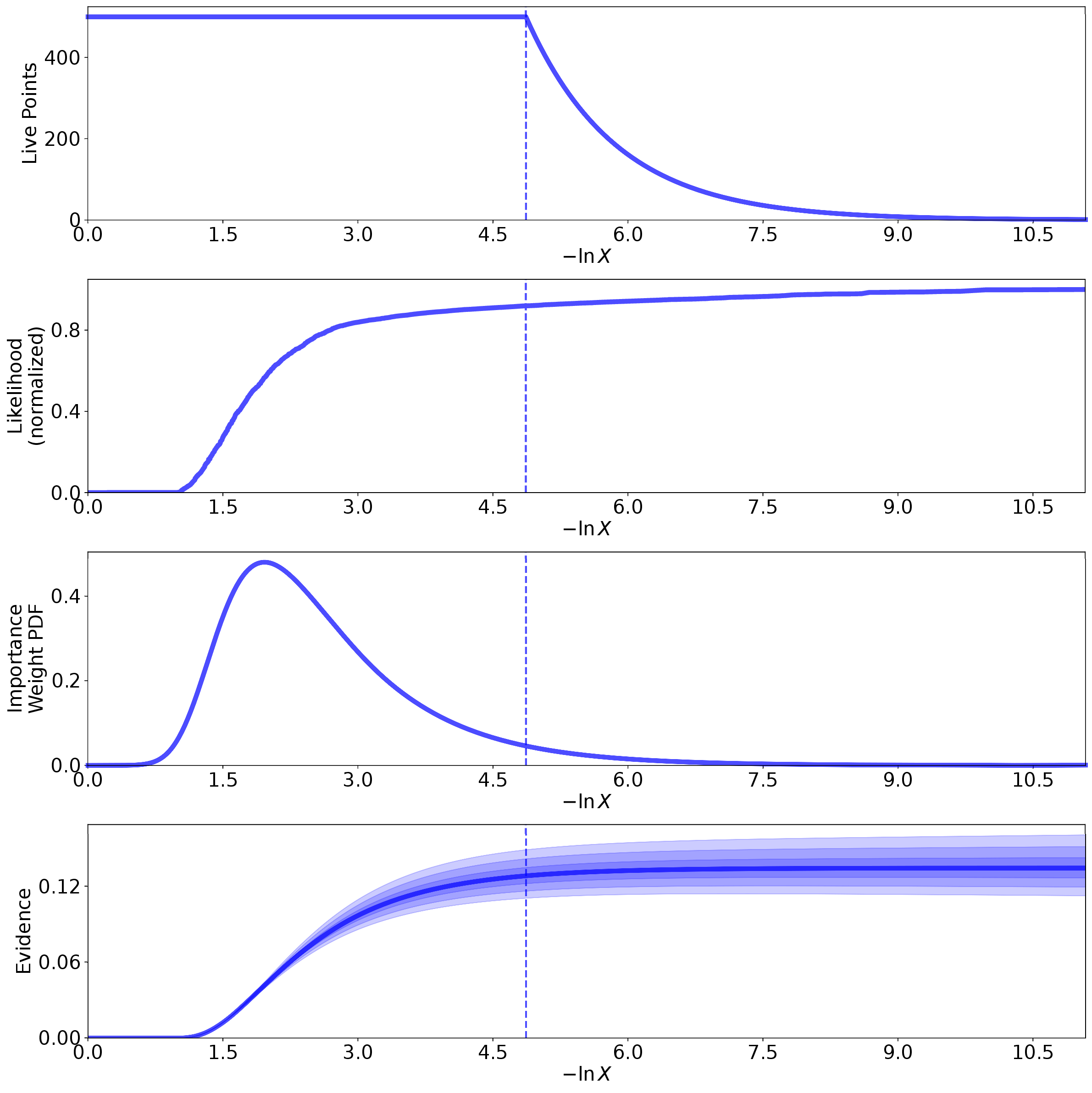}
		\caption{Statistical summary of the nested sampling for the \texttt{Linear}  equation of state model.}
		\label{stat_linear}
	\end{figure}

	The scale factor and Hubble rate evolution is represented for the \texttt{Linear} case against the standard observational limit in Fig.~\ref{aoft_linear}. We can observe that as this model allows for a varying equation of state parameter $w$, the dark energy fluid causes a quicker expansion of spacetime than the CC model, deviating from the standard trend at the 68\% bounds at the time of freezeout. At the 95\% upper limit, the initial density $\rho_{DE,0}$ is high enough to raise the scale factor above the standard expansion curve, as the Hubble rate decays toward a higher value than the standard framework. 
	
	We can verify that these two limits obtained for the parameters are numerically stable by studying the nested sampling diagnostics in Fig.~\ref{stat_linear}. The same qualitative result as in the \texttt{CC} model are indicating a clear prior volume exploration in the initial computing time, followed by a convergence towards a 95\% confidence level upper bound.
	
\begin{figure}[htbp!]
		\centering
		\includegraphics[scale =0.5]{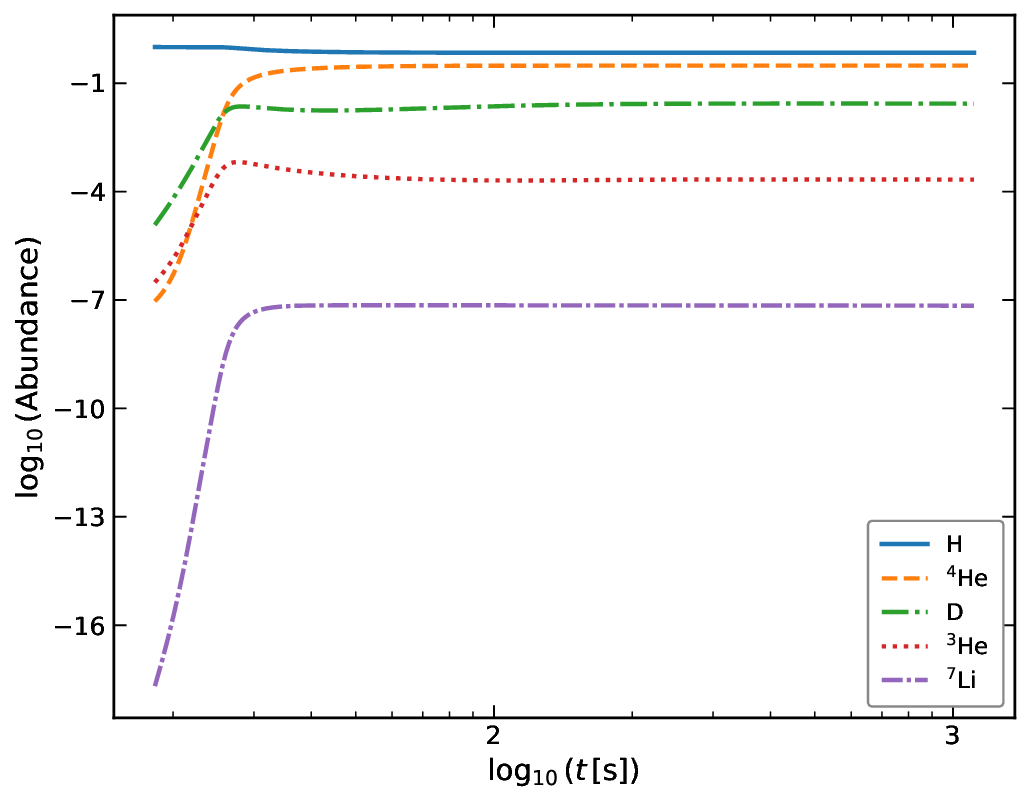}
		\caption{The log-log time evolution of light elements  for the \texttt{Linear} model.}
		\label{Yi_Lin}
	\end{figure}

	Figure~\ref{Yi_Lin} shows the abundances over time diagnostic for the \texttt{Linear} model. The evolutions of D, $^3$He and $^7$Li show no peaks and the synthesis timescale is shortened due to  the stronger coupling of $\rho_{\mathrm{EDE}}(a)$ to the expansion history in this model. For the light elements to form within the required bounds, a faster synthesis process is needed, which makes this model not recover the SBBN behaviour.

\begin{figure}[htbp!]
		\centering
		\includegraphics[scale =0.23]{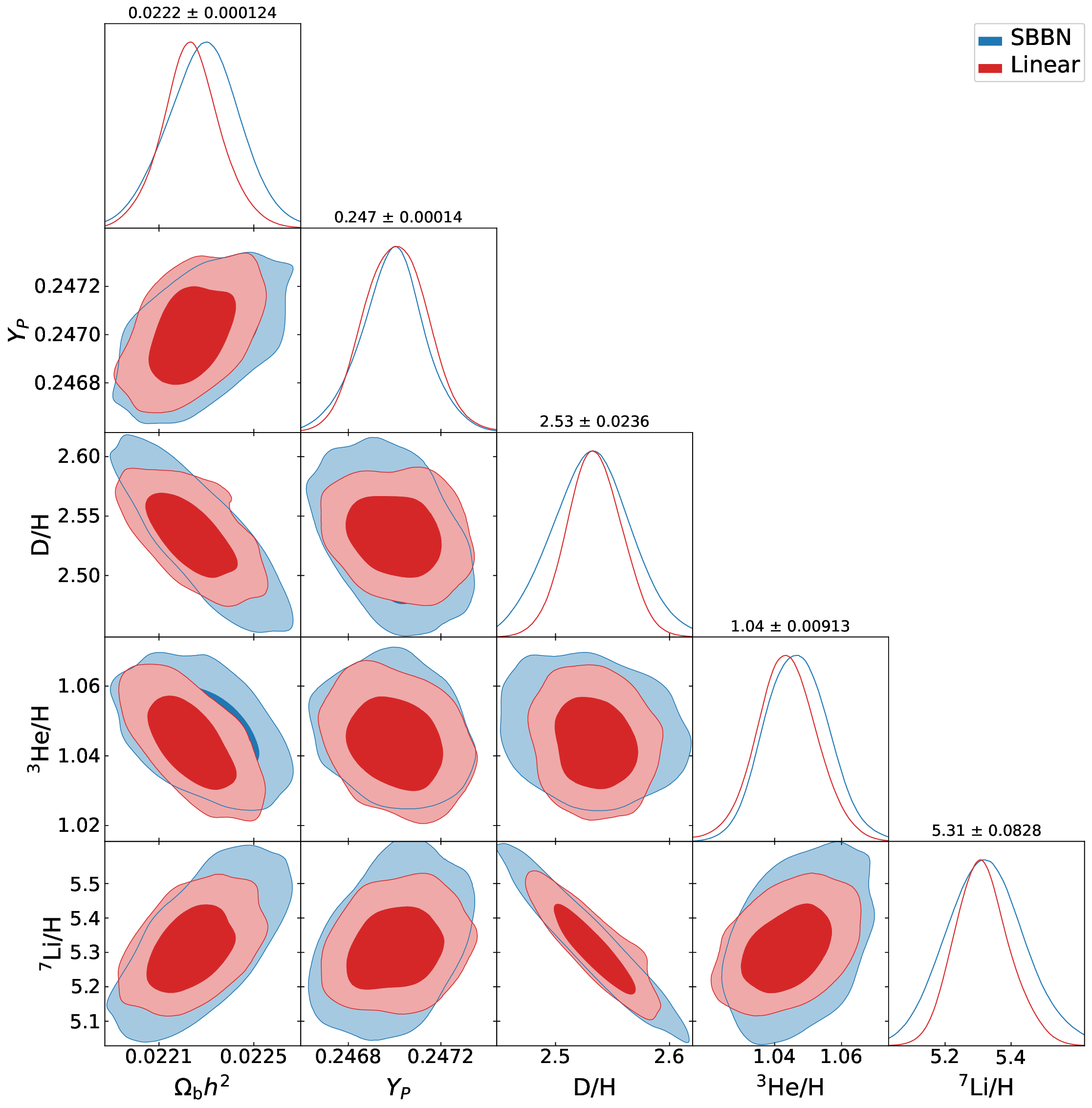}
		\caption{Nuclear abundances and $\Omega_b h^2$ for the \texttt{Linear} equation of state model overlaid with SBBN results from PRyMordial.}
		\label{abund_linear}
	\end{figure}

	Moreover, the abundance corner plots in Fig.~\ref{abund_linear} show the same	 strong overlap between the linear equation of state EDE model and the standard prediction, again an expected result since the upper bounds are indicating tiny contributions minimized through the $\chi$-squared function. 
	
	\subsection{The Polytropic EDE Model}
	
	In the \texttt{Polytropic} equation of state model, the dark energy fluid is parameterized to explore two physical behaviours, for which the polytropic index $\gamma = 4/3$ for a radiation-like component and $\gamma = 2$ for a stiff fluid. The fluid is assumed to scale as
	\begin{equation}
		\rho_{DE} = \left( \frac{a^{3(\gamma-1)}}{C} - K \right)^{\frac{1}{1-\gamma}},
	\end{equation}
	so that the behaviour of the fluid is determined by the competing effects of the expanding scale factor and the polytropic constant $K$. In the very early Universe, when the scale factor $a$ is vanishingly small, the polytropic pressure term dominates since $K\rho_{DE}^{\gamma} \gg \rho_{DE}$, and the condition $a^{3(\gamma-1)}/C \ll |K|$ is naturally satisfied,
	\begin{equation}
		\frac{a^{3(\gamma-1)}}{C} \ll |K| \quad \Rightarrow \quad \rho_{DE} \simeq (-K)^{\frac{1}{1-\gamma}},
	\end{equation}
	resembling a constant energy plateau that behaves like a cosmological constant. On the other hand, at late times the scale factor grows and the expansion term dominates, so that $K$ becomes negligible and the fluid reduces to pressureless dust diluting as,
	\begin{equation}
		\frac{a^{3(\gamma-1)}}{C} \gg |K| \quad \Rightarrow \quad \rho_{DE} \simeq C^{-1/(1-\gamma)}\, a^{-3}.
	\end{equation}
	To optimize numerical stability during nested sampling, we replace the integration constant $C$ with a scale factor $a_t$, defined at the transition between the dynamic and constant regimes. Moreover, for a real, positive energy density the polytropic constant is required to be strictly negative $K < 0$. The corresponding density plateau is $\rho_t = |K|^{1/(1-\gamma)}$, so that we can eliminate both $C$ and $K$ from the energy density relation, obtaining
	\begin{equation}
		\rho_{DE} = \rho_t \left[ \left( \frac{a}{a_t} \right)^{3(\gamma-1)} + 1 \right]^{\frac{1}{1-\gamma}}.
	\end{equation}
	For the above reparameterization we apply log-uniform priors in wide ranges,
	\begin{align}
		\log_{10} a_t &\sim \mathcal{U}(-15,\,-2), \\
		\log_{10} \rho_t &\sim \mathcal{U}(-20,\,5),
	\end{align}
	initializing the polytropic index with $4/3$ and $2$, respectively. 
	
	The nested sampling evaluation calculated a Bayesian evidence of $\ln \mathcal{Z}_{\mathrm{Poly}} = -1.5206 \pm 0.0538$ for the radiation-like fluid and $-1.5248 \pm 0.0539$ for the stiff fluid, with the resulting posterior constraints for the dynamic dark energy parameters detailed in Table \ref{poly_posteriors}.
	
	\begin{table}[h!]
		\centering
		\renewcommand{\arraystretch}{1.3}
		\begin{tabular}{lcc}
			\hline
			\text{Parameter} & \text{68\% CI} & \text{95\% CI} \\
			\hline
			\multicolumn{3}{c}{\texttt{Polytropic} ($\gamma = 4/3$)} \\
			\hline
			$a_t$ & $[2.17 \times 10^{-14},\, 1.05 \times 10^{-5}]$ & $1.44 \times 10^{-3}$ \\
			$\rho_t$ $[\mathrm{MeV}^4]$ &  $[7.63 \times 10^{-18},\, 2.52 \times 10^{-2}]$ & $2.75 \times 10^4$ \\
			\hline
			\multicolumn{3}{c}{\texttt{Polytropic} ($\gamma = 2$)} \\
			\hline
			$a_t$ &  $[1.96 \times 10^{-14},\, 2.19 \times 10^{-5}]$ & $1.42 \times 10^{-3}$ \\
			$\rho_t$ $[\mathrm{MeV}^4]$ &  $[1.05 \times 10^{-17},\, 1.76 \times 10^{-2}]$ & $8.13 \times 10^3$ \\
			\hline
		\end{tabular}
		\caption{Marginalized posterior constraints for the Polytropic  equation of state EDE models.}
		\label{poly_posteriors}
	\end{table}
	
\begin{figure}[htbp!]
		\centering
		\includegraphics[scale = 0.52]{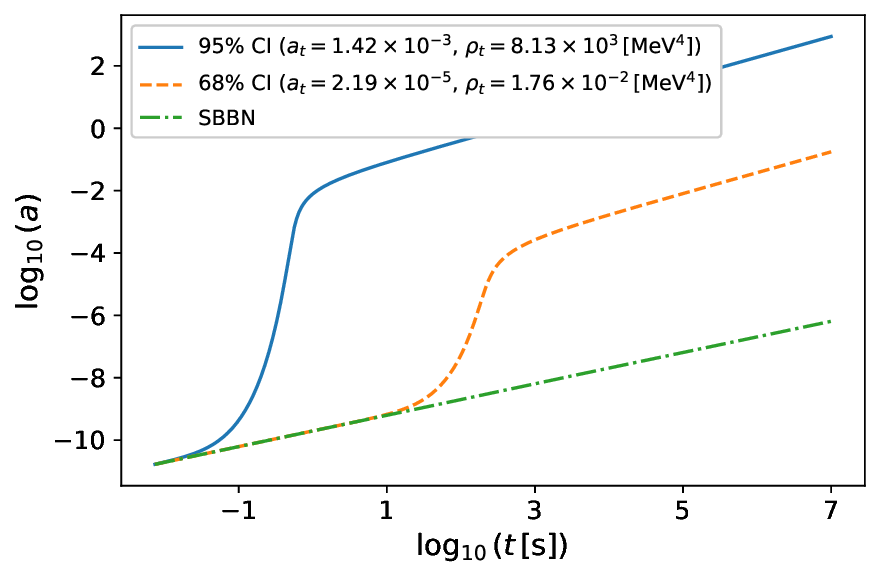}
		\includegraphics[scale = 0.52]{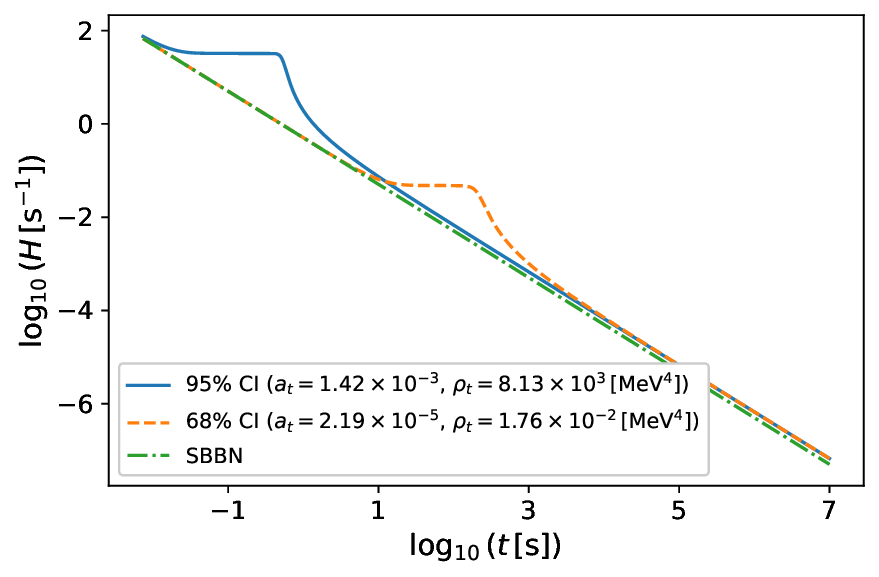}
		\caption{Evolution of the logarithm of the scale factor and the Hubble rate affected by the \texttt{Polytropic} equation of state for the representative radiation-like fluid ($\gamma = 4/3$) with parameters $a_{t}$ and $\rho_{t}$ at 95\% and 68\% CI compared to the SBBN trends.}
		\label{aoft_poly}
	\end{figure}

	The effects of the \texttt{Polytropic} dark energy fluid on the evolution of the scale factor and Hubble function are shown in Fig.~\ref{aoft_poly}. During the early-time regime where $a \ll a_t$, the dark energy density $\rho_t$ is constant and dominates the early dynamics causing a rapid evolution of the scale factor, pushing it significantly above the standard history (top panel). Following the transitory regime, the evolution trend is similar to the SBBN expansion in terms of the Hubble rate because both models reduce to pressureless dust diluting as $\rho_{DE} \propto a^{-3}$ at late times (bottom panel), whereas the scale factor achieves the largest value among all models. In both radiation and stiff-fluid scenarios, the dark component loses its pressure early in the evolution, meaning both quantities $\log a(t)$ and $\log H(t)$ show similar behaviours. Therefore, we present only the evolution plots for the $\gamma = 4/3$ radiation-like fluid as a representative example.
	
\begin{figure}[htbp!]
		\centering
		\includegraphics[scale=0.22]{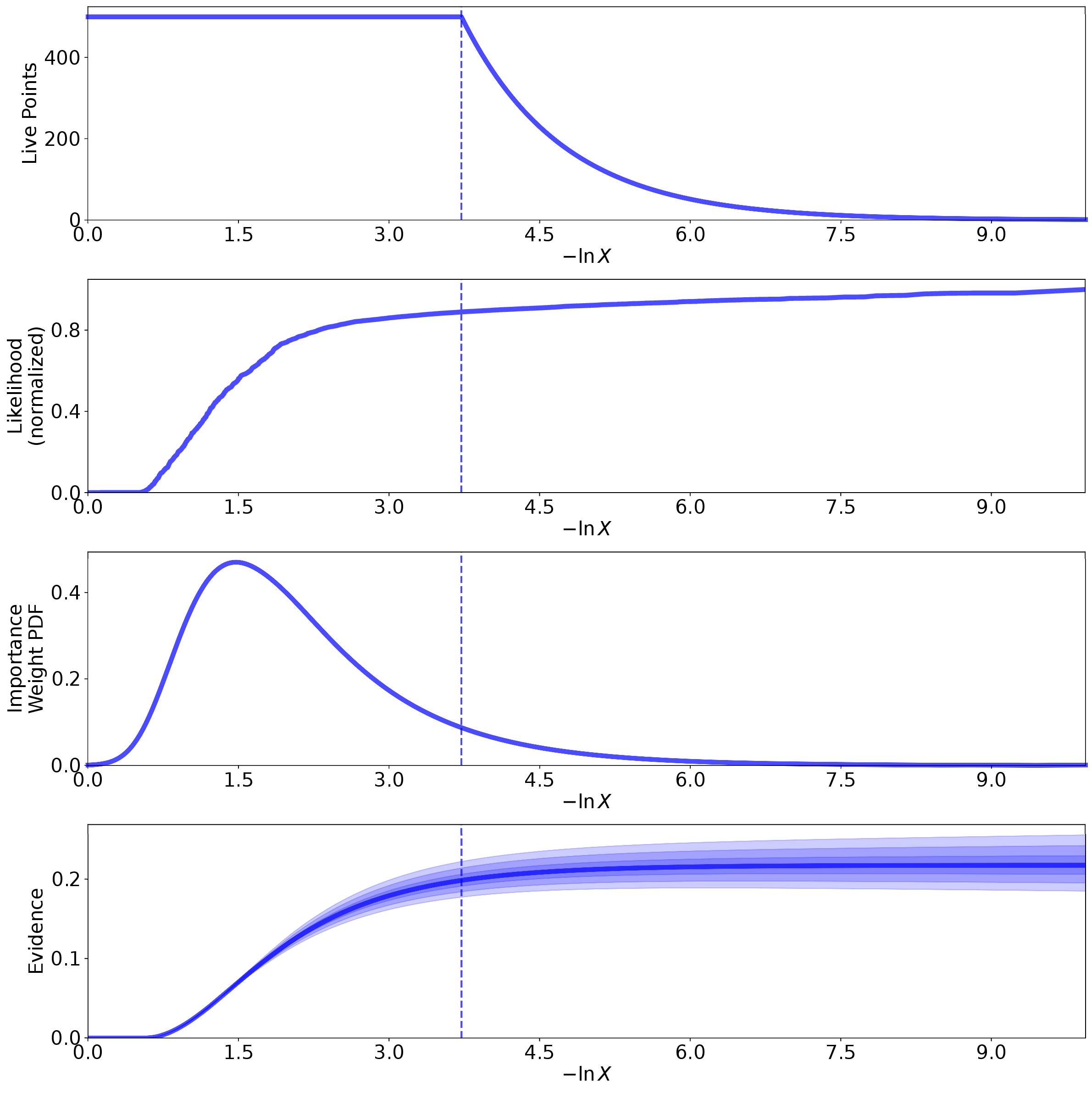}
		\caption{Statistical summary of the nested sampling for the \texttt{Polytropic}  equation of state model for the stiff fluid ($\gamma = 2$) polytropic index.}
		\label{stat_poly}
	\end{figure}

\begin{figure}[htbp!]
		\centering
		\includegraphics[scale =0.23]{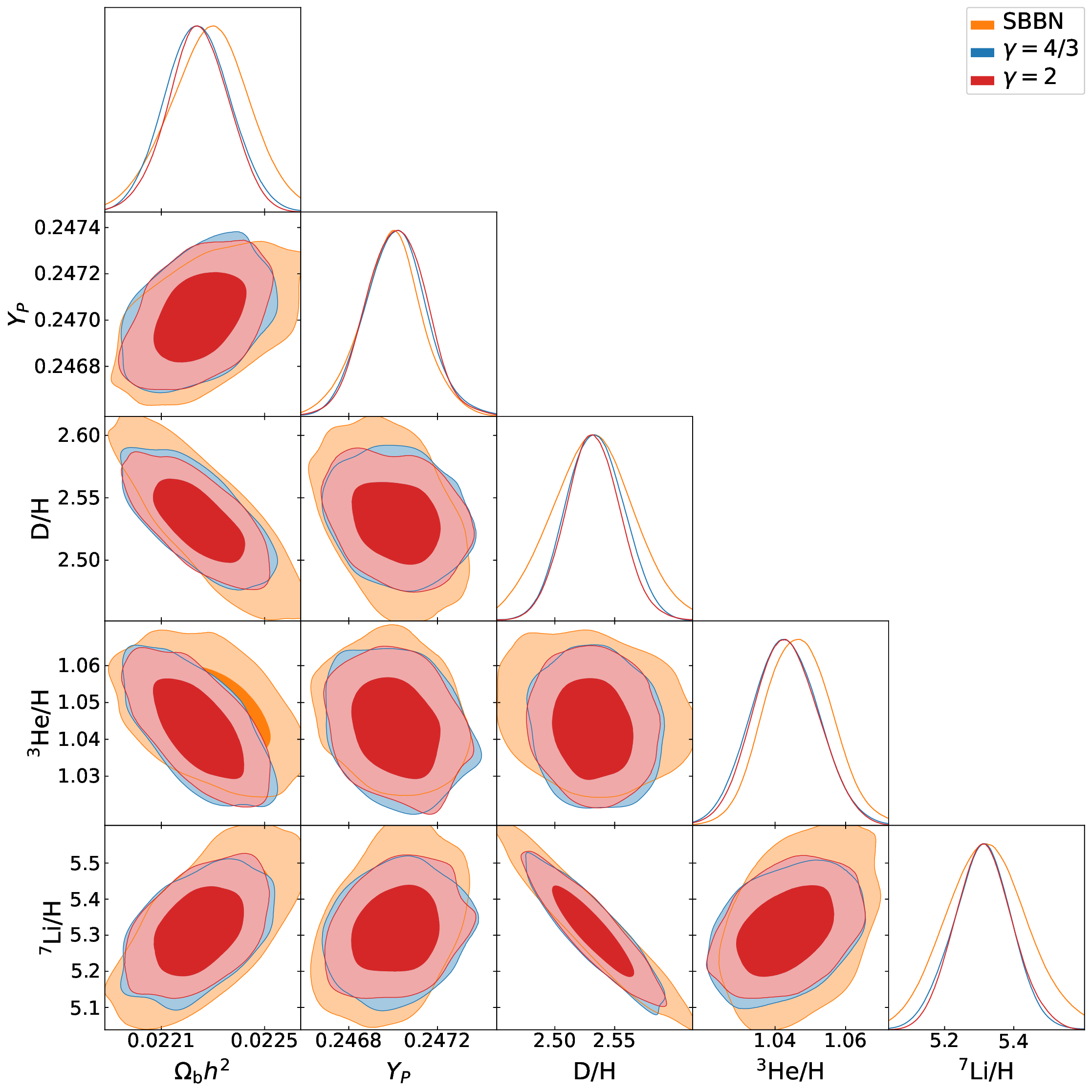}
		\caption{Nuclear abundances and $\Omega_b h^2$ for the \texttt{Polytropic} equation of state with both radiation and stiff fluid models overlaid with SBBN results from PRyMordial.}
		\label{abund_poly}
	\end{figure}

	The results from the two separate runs show similar transition parameters $a_t$ and $\rho_t$, namely for the radiation-like regime with 95\% upper limits of $a_t < 1.44 \times 10^{-3}$ and $\rho_t < 2.75 \times 10^4~\text{MeV}^4$, and for the stiff fluid with $a_t < 1.42 \times 10^{-3}$ and $\rho_t < 8.13 \times 10^3~\text{MeV}^4$, therefore we display in Fig.~\ref{stat_poly} only the nested sampling diagnostics for the $\gamma = 2$ model. The reason for the similarity in the model evidence, appears as we search for solutions close to the SBBN limit through our likelihood function, which is achieved for small enough contributions for the \texttt{Polytropic} model, so that the polytropic index doesn't modify significantly the overall final abundances, as it is seen in Fig.~\ref{abund_poly}.
	
\begin{figure}[htbp!]
		\centering
		\includegraphics[scale =0.5]{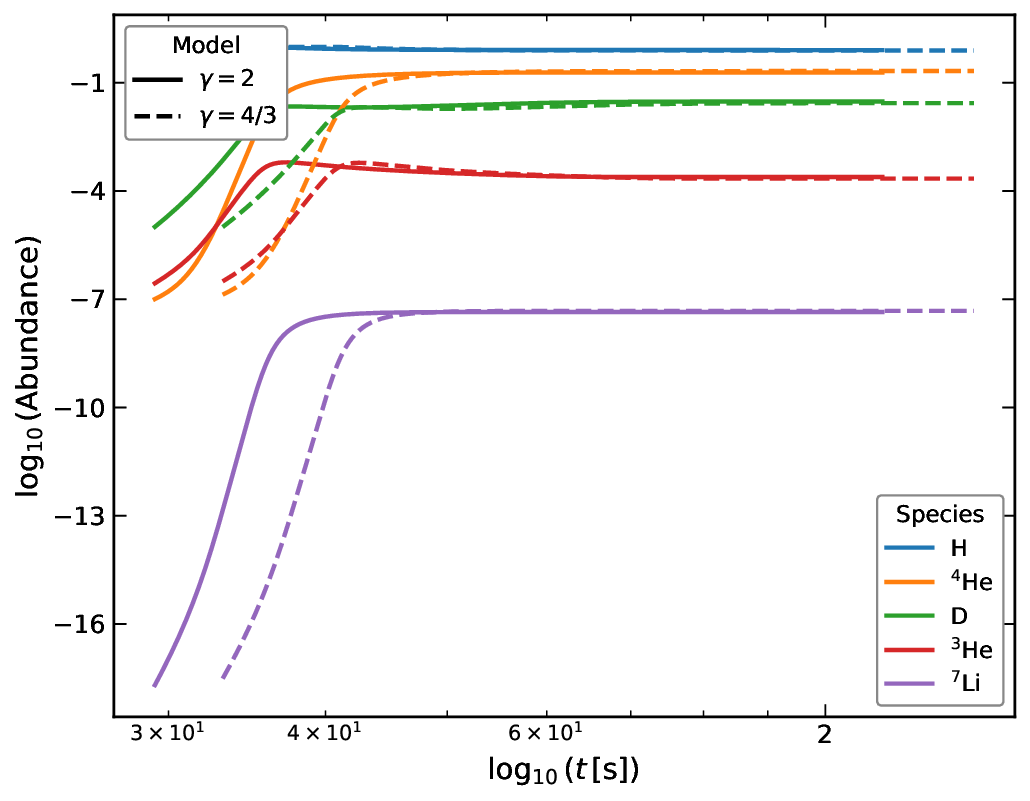}
		\caption{The log-log time evolution of light elements  for the two \texttt{Polytropic} models.}
		\label{Yi_Poly}
	\end{figure}

	The two \texttt{Polytropic} models are compared directly in terms of abundances evolution in Fig.~\ref{Yi_Poly}, where the solid and dashed curves correspond to $\gamma = 2$ and $\gamma = 4/3$, respectively. For these runs the abundances again freeze out after a very early shift of the nucleosynthesis epoch due to the strong $\rho_{\text{EDE}}(a)$ coupling to the background temperature, and the final relic levels are very similar, with an only difference between the two polytropes being the time to reach the maxima. Based on this analysis, we can conclude that a stiff fluid requires an earlier freezeout to recover the standard yields. Because the posterior constraints on the two polytropes are otherwise close, this time evolution is the only method to separate their BBN dynamics in the present analysis.

	\subsection{The Temperature-dependent EDE Model}
	
	For this last model, the equation of state for the dark energy component is varying with temperature, but still without any coupling assumed with the plasma sector. This allows us to write a linear temperature dependence of the equation of state parameter,
	\begin{align}
		\rho_{DE}(T) &= \rho_{DE,0} \left( \frac{T}{T_0} \right)^{3(1+w(T))}, \\
		w(T) &= -1 + \alpha T, 
	\end{align} 
	where $\alpha$ parameterizes the cosmological model.
	We initialize the prior volume by the uniform intervals, which is sufficiently low to allow for the existence of physical abundance ratios and provide upper limits,
	\begin{align}
		\log \rho_{T,0} &\sim \mathcal{U}(-40,\; -12), \\
		\alpha &\sim \mathcal{U}(0,\; 0.095),
	\end{align}
	
	The results of our simulations indicate that the model evidence $\ln \mathcal{Z}_{T} = -0.7774 \pm 0.0309$ is achieved for the upper limit parameters $\rho_{T,0} < 5.17\times 10^{-14}$ MeV$^4$ and $\alpha < 0.0886$, as shown in Table~\ref{LinearT}.
	
	\begin{table}[h!]
		\centering
		\renewcommand{\arraystretch}{1.3}
		\begin{tabular}{lcc}
			\hline
			\text{Parameter}  & \text{68\% CI} & \text{95\% CI} \\
			\hline
			$\rho_{DE,0}$ $[\mathrm{MeV}^4]$ & $[7.13 \times 10^{-36},\, 6.09 \times 10^{-17}]$ & $5.17 \times 10^{-14}$ \\
			$\alpha$ & $[1.31 \times 10^{-2},\, 7.77 \times 10^{-2}]$ & $8.86 \times 10^{-2}$ \\
			\hline
		\end{tabular}
		\caption{Marginalized posterior constraints for the temperature-dependent  equation of state model.}
		\label{LinearT}
	\end{table}
	
\begin{figure}[htbp!]
		\centering
		\includegraphics[scale = 0.52]{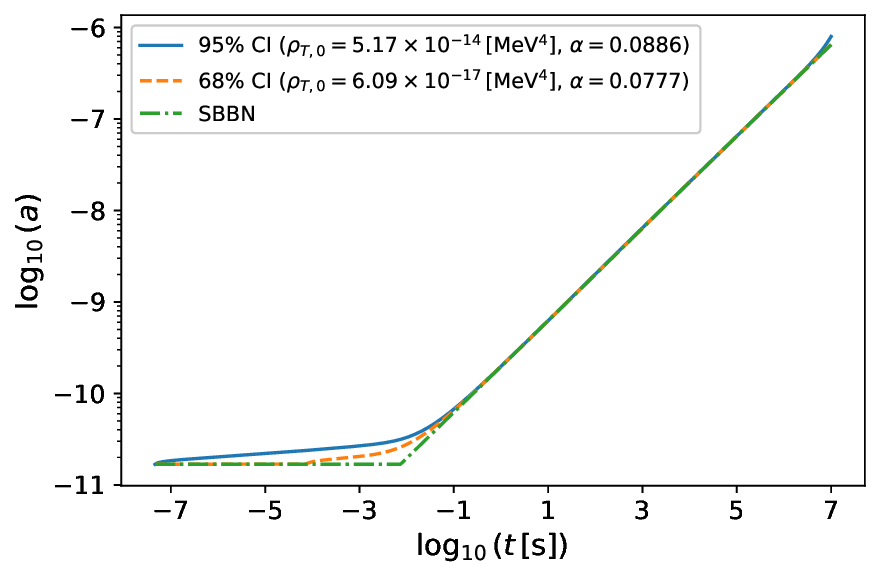}
		\includegraphics[scale = 0.52]{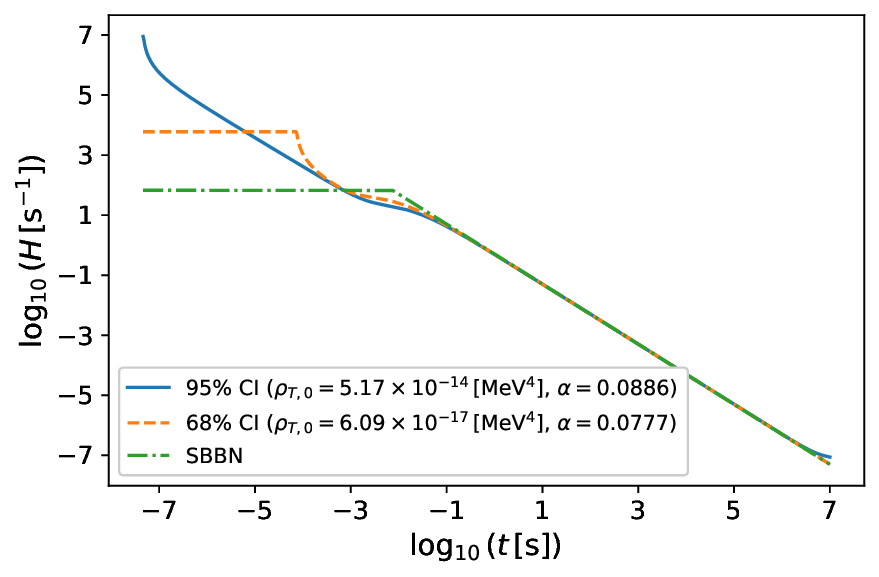}
		\caption{Evolution of the logarithm of the scale factor and the Hubble rate affected by the temperature-dependent equation of state for the \texttt{Linear w(T)} model, at 95\% and 68\% CI compared to the SBBN trends.}
		\label{aoft_tdep}
	\end{figure}

\begin{figure}[htbp!]
		\centering
		\includegraphics[scale=0.22]{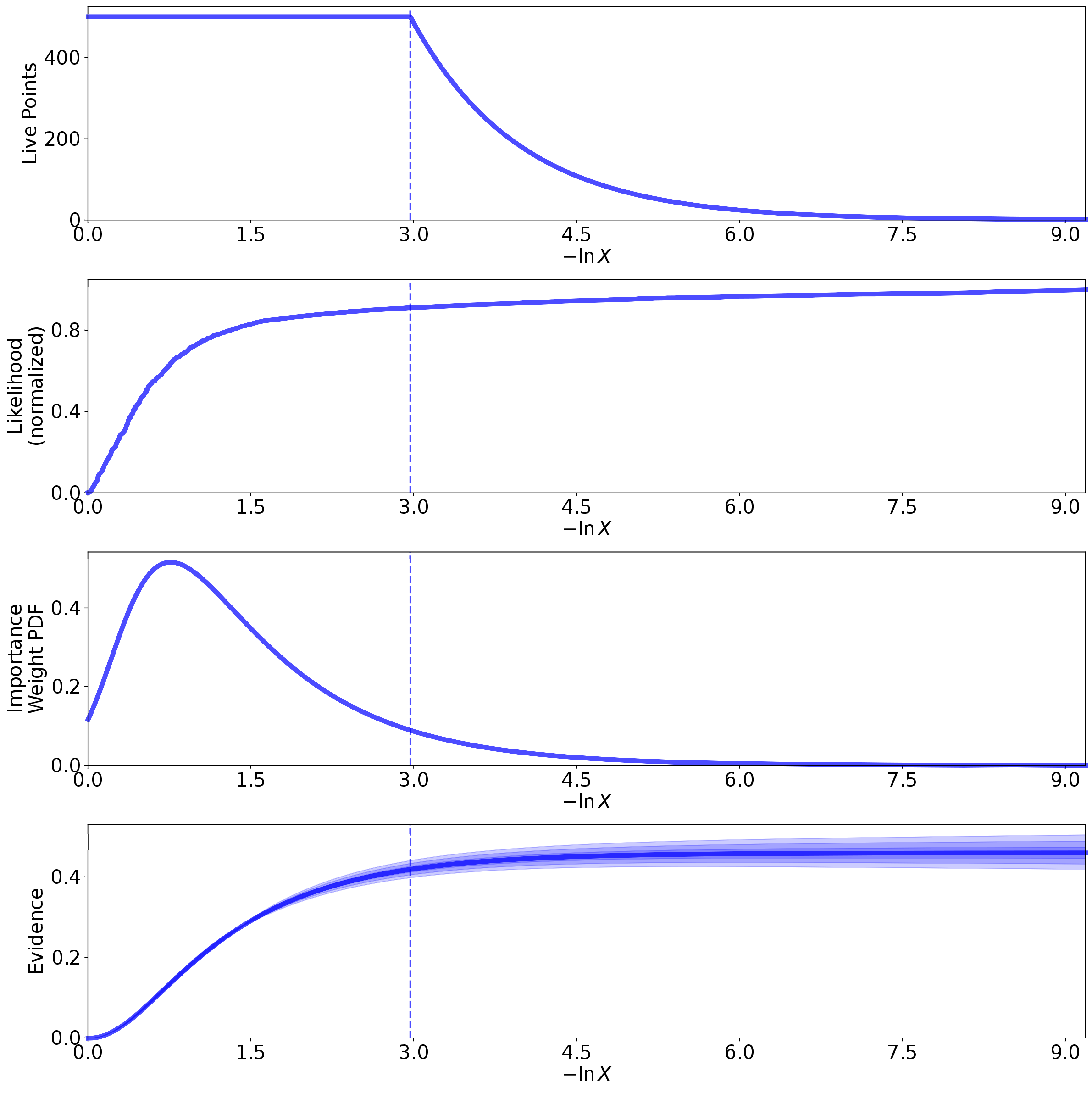}
		\caption{Statistical summary of the nested sampling for the \texttt{Linear w(T)}  equation of state model with temperature dependence.}
		\label{stat_tdep}
	\end{figure}

\begin{figure}[htbp!]
		\centering
		\includegraphics[scale =0.23]{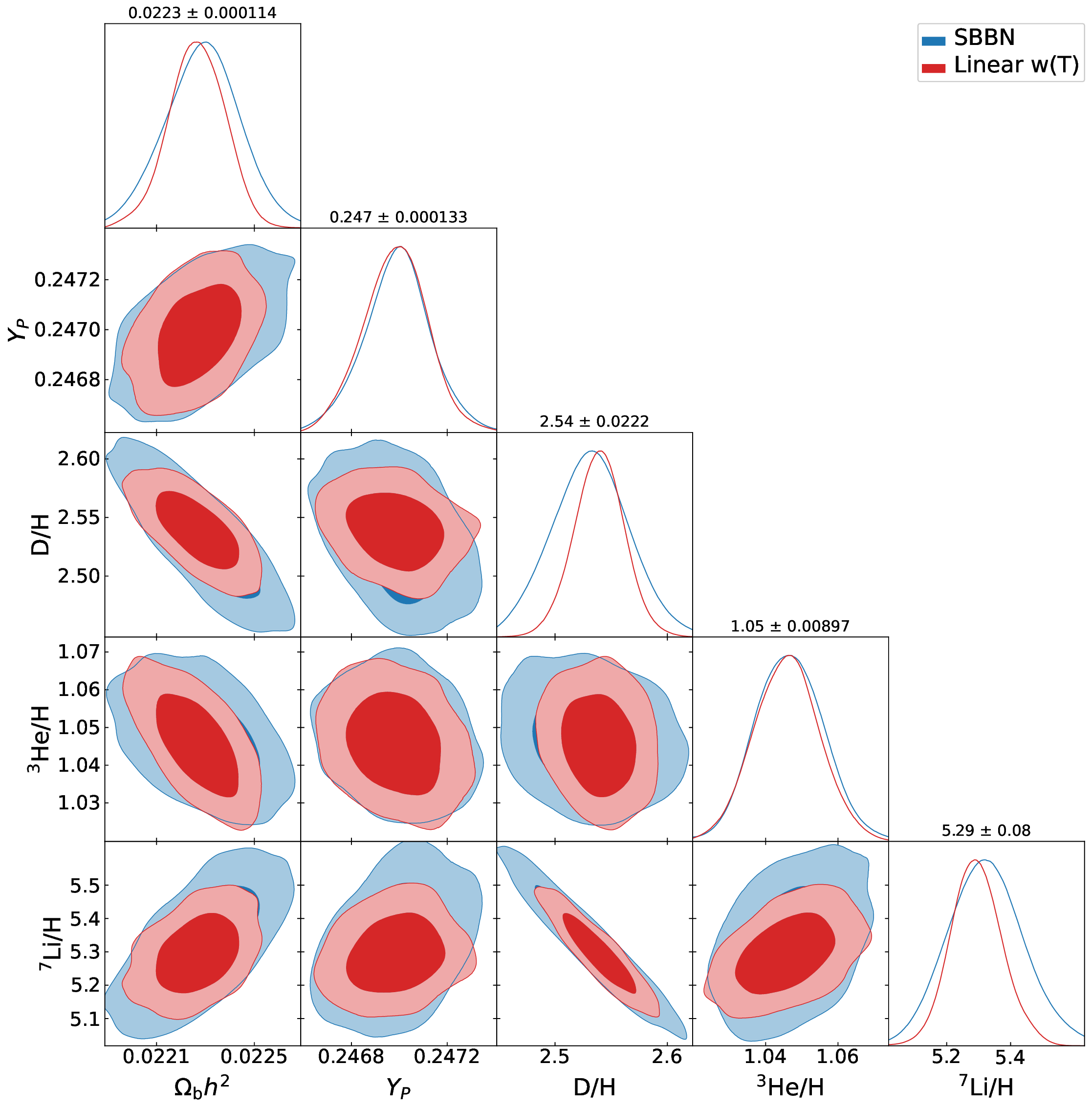}
		\caption{Nuclear abundances and $\Omega_b h^2$ for the temperature-dependent equation of state compared with the SBBN results from PRyMordial.}
		\label{abund_tdep}
	\end{figure}

	The logarithmic plots in Fig~\ref{aoft_tdep} show the smallest deviation from the SBBN model as the scale factor doesn't increase at scales grater than the standard results, as opposed to the other time-dependent/constant models in which the expansion scale increased due to the modified background dynamics. In terms of the logarithmic timescale, this model emphasises a greater contribution of the dark sector at the very begining of the simulation, as the energy density is directly proportional to the temperature. This causes the ODE solver to prioritize the very early time domain in the integration, where deviations from the SBBN model are observed, which eventually fade out as the Universe cools. Therefore, the contribution of this particular model to the nucleosynthesis process fades as temperature drops, so that all abundance posteriors are well within the SBBN bounds, visible in Fig.~\ref{abund_tdep}. The summary of the model convergence shows a similar trend as for the other investigated models, displayed in Fig.~\ref{stat_tdep}.
	
\begin{figure}[htbp!]
		\centering
		\includegraphics[scale =0.5]{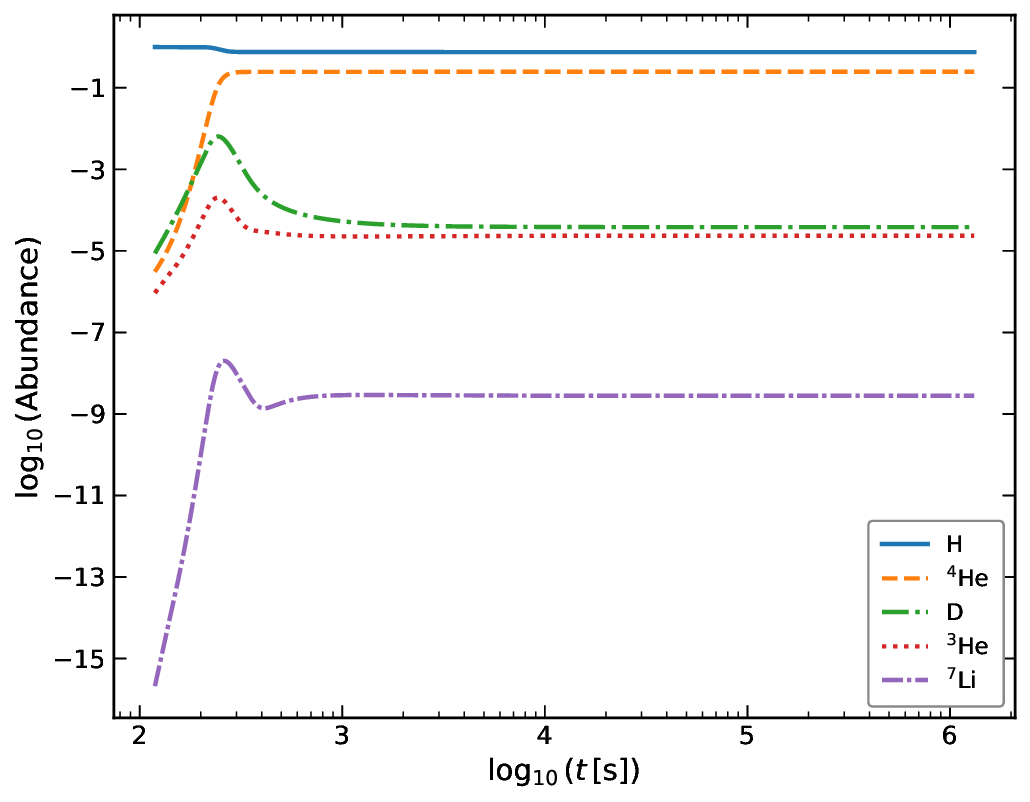}
		\caption{The log-log time evolution of light elements  for the \texttt{LinearT} model.}
		\label{Yi_Tdep}
	\end{figure}

	Finally, Fig.~\ref{Yi_Tdep} displays the temperature-dependent model and the \texttt{CC} case are the only scenarios that recover the standard BBN timescale and the characteristic maxima in abundances, as opposed to the \texttt{Linear} and \texttt{Polytropic} models, which shift nucleosynthesis to earlier times.

	\section{Discussions and Final Remarks} \label{final}
	
	In the present work, we studied the influence of a constant and various dynamical Early Dark Energy models on the nucleosynthesis process which takes place in the very early Universe. We incorporated the effects of a cosmological constant model, one linear, and two polytropic time-dependent equation of state models, for which a radiation-like dark component and a stiff fluid were considered. We also studied the temperature evolution for a linear equation of state model, by including these modifications into the background equations of \texttt{PRyMordial}'s modeling of the first Friedmann equation. By using a nested sampling technique, we extracted robust posterior constraints and upper limits on the energy densities of these dark sector, establishing the maximum physical values allowed before violating observed primordial light element abundances. 
	
	In what follows, we evaluate the proposed Early Dark Energy models by comparing their statistical performance against both the standard Big Bang Nucleosynthesis predictions and each other. The evidence extracted from the nested sampling algorithm provides a sufficiently robust metric for model selection, hence we establish the relative preference between models using the Bayes Factor, expressed logarithmically as $\Delta \ln \mathcal{Z} = \ln \mathcal{Z}_{\text{model}} - \ln \mathcal{Z}_{\text{CC}}$, where we used the Cosmological Constant model as our baseline as it is the simplest one in terms of the number of degrees of freedom. According to the Kass and Raftery scale \cite{KassRaftery}, a difference $|\Delta \ln \mathcal{Z}| < 1$ indicates that the models are statistically indistinguishable, whereas values greater than 3 would suggest strong evidence in favour of one model over the other. The statistical properties of our EDE models are summarized in Table~\ref{stat_summary}, and show nearly equal statistical preference.
	
	\begin{table}[h!]
		\centering
		\renewcommand{\arraystretch}{1.3}
		\begin{tabular}{lcc}
			\hline
			\text{Model} & $\ln \mathcal{Z}$ & $\Delta \ln \mathcal{Z}$ \\
			\hline
			\texttt{Cosmological Constant} ($\Lambda$) & $-1.290 \pm 0.044$ & $0.000$  \\
			\texttt{Linear} ($w = \text{const.}$) & $-2.006 \pm 0.059$ & $-0.716$  \\
			\texttt{Polytropic} ($\gamma = 4/3$) & $-1.521 \pm 0.054$ & $-0.231$  \\
			\texttt{Polytropic} ($\gamma = 2$) & $-1.525 \pm 0.054$ & $-0.235$ \\
			\texttt{Linear w(T)} ($w(T) = -1 + \alpha T$) & $-0.774 \pm 0.031$ & $+0.516$ \\
			\hline
		\end{tabular}
		\caption{Statistical comparison of the EDE models through Bayes factors $\Delta \ln \mathcal{Z}$ calculated relative to the cosmological constant model.}
		\label{stat_summary}
	\end{table}
	 
	We can visualize the upper limits obtained in terms of the physical energy scales allowed during the BBN epoch. In the present-day Universe, observations of the cosmological constant are constrained to $\rho_{\Lambda, 0} \approx 10^{-56}$~cm$^{-2}$, and we can place our early-Universe results in this context by comparing our results in geometric density units, as summarized in Table~\ref{energy_density}. It is important to note that in the \texttt{Polytropic} models, the parameter $\rho_t$ represents a transient high-redshift density plateau and in order to compare it against present-day density parameters we must account for the fluid's dilution in the late-time limit. Because both models analytically reduce to an effective $\rho_{DE} \simeq \rho_t (a/a_t)^{-3} = \rho_t a_t^3 a^{-3}$, the equivalent present-day density parameter is expressed as the scaled ratio $\rho_t a_t^3$ in Table~\ref{energy_density}. Even so, the allowed energy density for the polytropic models remains several orders of magnitude larger than the strict limits placed by the \texttt{Linear} or \texttt{Cosmological Constant} models.
	
	\begin{table}[h!]
		\centering
		\renewcommand{\arraystretch}{1.3}
		\begin{tabular}{llc}
			\hline
			Model & Density Parameter [MeV$^4$] &  [cm$^{-2}$] \\
			\hline
			\texttt{CC} & $\Lambda < 9.41 \times 10^{-12}$ & $4.07 \times 10^{-33}$ \\
			\texttt{Linear} & $\rho_{DE,0} < 2.45 \times 10^{-13}$ & $1.06 \times 10^{-34}$ \\
			\texttt{Polytropic} ($\gamma=\frac{4}{3}$) & $\rho_t a_t^3 < 8.21 \times 10^{-5}$ & $3.55 \times 10^{-26}$ \\
			\texttt{Polytropic} ($\gamma=2$) & $\rho_t a_t^3 < 2.33 \times 10^{-5}$ & $1.01 \times 10^{-26}$ \\
			\texttt{Linear w(T)} & $\rho_{T,0} < 5.17 \times 10^{-14}$ & $2.24 \times 10^{-35}$ \\
			\hline
		\end{tabular}
		\caption{Upper limits for the energy density scale at 95\% CI across the tested EDE models. To ensure a consistent baseline comparison, the polytropic models are expressed via their equivalent present-day density.}
		\label{energy_density}
	\end{table}
	
	Even though the EDE models cannot be ranked in terms of Bayesian statistical metrics, the physical viability of these models can be understood based on the kinematic evolution of the Hubble expansion rate, $H(t)$, which governs the decoupling of weak interactions and the neutron-to-proton ratio freeze-out at $T \sim 1$ MeV. 

The cosmological constant model represents the most rigid scenario, being characterized by a non-diluting energy density which interrupts the natural $t^{-1/2}$ decay of the Hubble parameter during radiation domination era prior to recombination. On the other hand, a dynamic fluid with a linear, time-dependent equation of state dilutes as $\rho_{DE} \propto a^{-3(1+w)}$, maintaining a slightly higher Hubble rate relative to the SBBN background throughout nucleosynthesis.
	
	The polytropic models introduce a different kinematic evolution as they asymptotically reduce to a pressureless dust-like fluid at late times. Their resulting physical behaviours during nucleosynthesis are practically indistinguishable, causing the 95\% upper limit Hubble rate shows slight deviations from the standard BBN curve during the weak freeze-out epoch compared to the corresponding stagnation of the $H(t)$ rate experienced at 68\% limit. Finally, the temperature-dependent linear model $w(T)$ modifies the background expansion strictly at high temperatures, but the evolution curves remain identical to the standard SBBN profile, making it hardly noticeable in the abundance ratios.
	
	The EDE contribution to the background dynamics indirectly affects the neutrino decoupling mechanism by altering the total effective number of relativistic species $N_{\text{eff}}$ through the modified $H(t)$ evolution, which governs the coupling between the photon plasma and the three neutrino species. Using the 95\% upper limit EDE parameters, the CC and temperature-dependent models, whose EDE energy density does not depend on the scale factor evolution, recover the standard $N_{\text{eff}} \simeq 3.044$. The linear and polytropic models, however, require the scale factor to be recomputed in the background evolution, which introduces a tighter coupling between $\rho_{\text{EDE}}$ and the expansion rate during neutrino decoupling, resulting in $N_{\text{eff, Linear}} \simeq 3.010$, $N_{\text{eff, Poly}}(\gamma=4/3) \simeq 3.017$ and $N_{\text{eff, Poly}}(\gamma = 2) \simeq 3.015$. The average shift $\Delta N_{\text{eff}} \equiv N_{\text{eff}} - 3.044 \simeq -0.03$ remains well within observational bounds, given that current BBN and CMB data constrain $N_{\text{eff}} = 2.898 \pm 0.141$ at the $2\sigma$ level \cite{Yeh2022}, confirming that the EDE contribution to the expansion rate during neutrino decoupling does not introduce tension with the measured radiation content at BBN.
	
\begin{figure}[htbp!]
		\centering
		\includegraphics[scale=0.5]{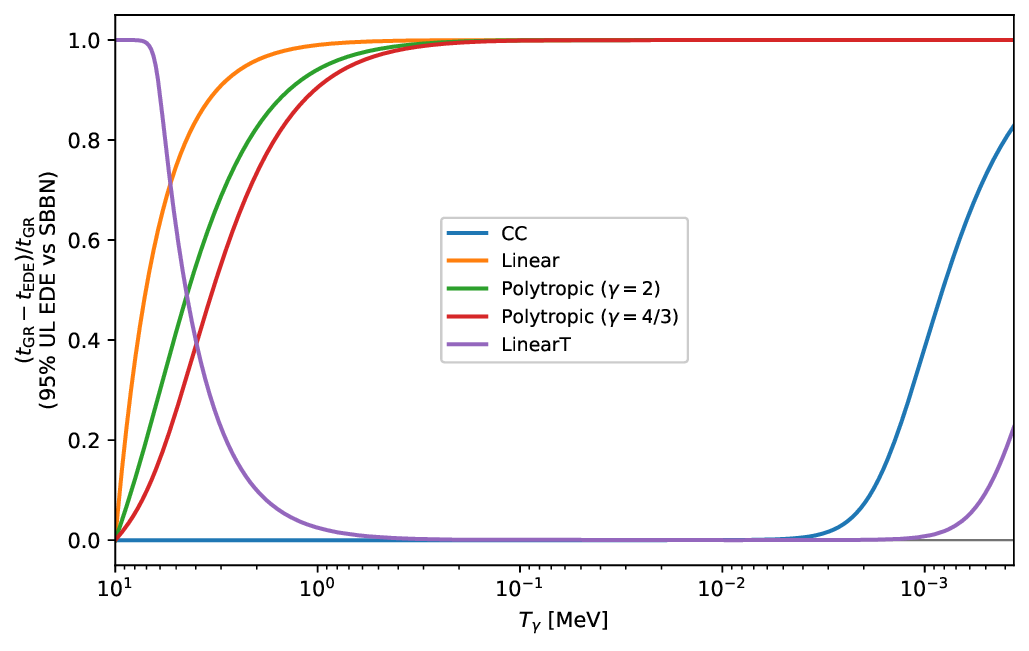}
		\caption{Deviations of the relative time difference of GR and EDE models as a function of photon temperature for each cosmological model considered.}
		\label{t_of_T}
	\end{figure}

	To quantify the divergence from General Relativity, we define the fractional difference in cosmic time evaluated for the photon temperature $T_\gamma$ as $(t_{\mathrm{GR}} - t_{\mathrm{EDE}}) / t_{\mathrm{GR}}$. We perform a model parameter evaluation at the 95\% upper limit for each EDE model, and we make the distinction between a positive cosmic time ratio, which indicates the EDE universe reaches a specific temperature sooner than the GR limit, and a negative ratio, which conversely, signals that the expansion history is behind the GR baseline. By plotting this as a function of $T_\gamma$ from the hot BBN onset at $10$ MeV toward lower values, we display the effect of the energy density of the dark sector on the time evolution in Fig.~\ref{t_of_T}.
	
	The results show that all EDE models, besides the temperature-dependent equation of state, start from a zero deviation from the GR limit at high temperatures, which indicates that the EDE effects contribute gradually to the total energy density of the background as time passes. Only the \texttt{LinearT} model representing the temperature-dependent equation of state starts with a maximum value associated with the highest temperature, which then rapidly dilutes as the Universe cools down. 

For the \texttt{Linear} and \texttt{Polytropic} equation of state of the dark components, the relative cosmic time difference increases sharply, so that by the time of weak freeze-out at $T_\gamma \approx 0.5$ MeV, the relative difference stabilizes at the maximally allowed deviation. This is an expected behaviour reflecting the cumulative temporal shift evaluated at our 95\% upper limit constraints. 

For the \texttt{CC} model, we observe a rapid increase in time deviation from GR only at much lower temperatures, $T_\gamma \lesssim 10^{-2}$ MeV, approaching the nucleosynthesis regime. This late-time deviation occurs because the cosmological constant does not dilute, hence as the background radiation energy density drops proportionally to $T^4$, the constant \texttt{CC} energy density eventually becomes a significant fraction of the total energy of the Universe, modifying the expansion rate. Similarly, a secondary relative increase happens in the temperature-based \texttt{LinearT} model, where at low temperatures, the energy density stops from diluting as rapidly as the Universe cools, causing it to behave as a cosmological constant at later times.
	
	Although SBBN predictions are in good agreement with primordial abundance observations, EDE models are frequently used to resolve the $H_0$ tension, which would require a higher early-time Hubble parameter to reduce the physical comoving sound horizon at recombination, 
	\begin{equation}
		r_s = \int_{z_{rec}}^{\infty} \frac{c_s(z)}{H(z)} dz. 
	\end{equation}	

	It should be noted that the scope of the present work is strictly limited to the nucleosynthesis epoch and the resulting primordial abundances as simulated by the \texttt{PRyMordial} framework. While our results provide a necessary validation for early dark energy sectors motivated by the Hubble tension, we do not explicitly integrate the sound horizon $r_s$, or perform a full cosmic microwave background likelihood analysis. Instead, we establish the maximum allowed limits these fluids can reach without violating current nuclei data.
	
	Finally, although our analysis targeted only early-time nucleosynthesis data and lacks a late-time constraining solution, the viability of an Early Dark Energy model depends on its capacity to balance both epochs. From this qualitative perspective, the temperature-dependent equation of state \texttt{LinearT} emerges as the best candidate among our models, as it allows for significant modifications to the expansion history at extreme temperatures, while rapidly diluting its energy density to align with the General Relativity limit during the weak freeze-out era $T_\gamma \approx 0.5$~MeV. Importantly, among all models, the \texttt{LinearT} and \texttt{CC} are the only scenarios that correctly predict the BBN timescale and evolution for nucleosynthesis.

Supported by a slight statistical preference in our Bayesian selection $\Delta \ln \mathcal{Z} = +0.516$ and its theoretical ability to re-emerge as a cosmological constant at very low temperatures, the temperature-dependent model manages to efficiently satisfy the strict bounds of primordial element formation, which makes it a physical scenario worth studying in future works.

\section{Acknowledgments}

We would like to thank the anonymous referee for comments and suggestions that helped us to significantly improve our work.

\end{document}